# CHAPTER 2

# A PORTFOLIO REBALANCING APPROACH FOR THE INDIAN STOCK MARKET

JAYDIP SEN, ARUP DASGUPTA, SUBHASIS DASGUPTA & SAYANTANI ROY CHOUDHURY

## Introduction

The process of managing a portfolio must include portfolio rebalancing. It offers security and control for any professional or retail investment management plan. First and foremost, portfolio rebalancing protects the investor from being overexposed to risky situations. Additionally, it makes sure that the exposures in the portfolio stay within the manager's competence. Let us consider an investor who has invested 75% of their portfolio in risk-free assets and the remaining 25% in stocks. If the equity assets treble in value, then risky stocks now make up 50% of the portfolio. Given that the allocation has changed and is now outside of their area of competence, a portfolio manager who is only competent to manage fixed-income assets would no longer be able to manage the portfolio. The portfolio must be rebalanced often to prevent these undesirable movements. Additionally, the rising percentage of the portfolio that is invested in stocks raises the total risk to levels above what an investor would typically choose.

There are three well-known portfolio rebalancing techniques (i) calendar rebalancing, (ii) percentage-of-portfolio rebalancing, and (iii) constant proportion portfolio insurance.

*Calendar Rebalancing:* In this approach, rebalancing is done based on the calendar. This method only entails reviewing the portfolio's investment holdings at preset intervals and making necessary adjustments to return to the original allocation. Since weekly rebalancing would be too costly and a yearlong method would allow for too much intermediate portfolio drift, monthly and quarterly reviews are often recommended. Time limits, transaction costs, and permitted drift must all be considered while determining the optimal rebalancing frequency. A portfolio with a 60/40 split between stocks and bonds that was rebalanced monthly, quarterly, yearly, or never was the subject of a 2019 Vanguard research (Pagliaro & Utkas, 2019). Jaconetti et al. (2010), between the various time frames, discovered "little difference" in portfolio performance. In comparison to other formula-based rebalancing techniques, calendar rebalancing is substantially less time-consuming for investors because it is a continuous operation.

*Percentage-of-Portfolio Rebalancing*: Rebalancing based on the permissible percentage composition of an asset in a portfolio is a preferred but marginally more time-consuming way to put it into practice. There is a goal weight and a matching tolerance range assigned to each asset class or individual security. For instance, an allocation plan may mandate holding 40% of government bonds, 30% of domestic blue chips, and 30% of developing market stocks, with a +/- 5% range for each asset class. Stakes in emerging markets and domestic blue-chip companies might vary between 25% and 35%, while government bonds must make up 35% to 45% of the portfolio. The whole portfolio is rebalanced to match the initial target composition when the weight of any holding crosses the permissible band.

*Constant Proportion Portfolio Insurance* (CPPI): This strategy of portfolio rebalancing assumes that as investors' wealth rises, so does their risk tolerance. The fundamental idea behind this approach is that it is preferable to keep a minimal safety reserve stored in either cash or risk-free government bonds. Consequently, this strategy entails constantly altering the allocation between risky and risk-free assets following market conditions. More money is invested in stocks as the value of the portfolio rises, whereas a decline in portfolio value results in a lower position in risky assets. The most crucial necessity for the investor is to maintain the safety reserve, regardless of whether it will be utilized to pay for college expenses or a down payment on a property. As CPPI rebalancing does not specify the frequency of rebalancing and only specifies how much equity should be held in a portfolio, it must be used in conjunction with rebalancing and portfolio optimization strategies. Additionally, it does not provide a holding breakdown of asset classes along with their ideal corridors.

Portfolio rebalancing reduces risk by avoiding overexposing investors to volatile assets over the long term, but it comes at a cost. Taxes and transaction fees are the two primary expenses to take into account while rebalancing a portfolio. Fees from fund managers, for instance, may be associated with each rebalancing transaction. Sales of assets may result in capital gains or losses that affect taxes.

This chapter presents a calendar rebalancing approach to portfolios of stocks in the Indian stock market. Ten important sectors of the Indian economy are first selected. For each of these sectors, the top ten stocks are identified based on their free-float market capitalization values. Using the ten stocks in each sector, a sector-specific portfolio is designed. In this study, the historical stock prices are used from January 4, 2021, to September 20, 2023 (NSE Website). The portfolios are designed based on the training data from January 4, 2021 to June 30, 2022. The performances of the portfolios are tested over the period from July 1, 2022, to September 20, 2023. The calendar rebalancing approach presented in the chapter is based on a yearly rebalancing method. However, the method presented is perfectly flexible and can be adapted for weekly or monthly rebalancing.  The rebalanced portfolios for the ten sectors are analyzed in detail for their performances. The performance results are not only indicative of the relative performances of the sectors over the training (i.e., in-sample) data and test (out-of-sample) data, but they also reflect the overall effectiveness of the proposed portfolio rebalancing approach.

The work has three unique contributions. First, it presents a rebalancing approach for stock portfolios which can be adapted at yearly, monthly, and daily levels. Second, the portfolios are backtested using several metrics including, cumulative returns, annual volatilities, and Sharpe ratios. The results of the evaluation identify the best-performing portfolio corresponding to each sector of stocks over the training and the test periods. Finally, the results of this study provide a deep insight into the current profitability of the sectors that will be useful for investors in the Indian stock market.

The chapter is organized as follows. The section titled *Related Work* presents some of the existing portfolio design approaches proposed in the literature. Next, the section titled *Methodology* presents the research approach followed in the current work. The section titled *Results* presents an extensive set of results and a detailed analysis of the observations. Finally, the chapter is concluded in the section titled *Conclusion*.

# Related Work

Designing and optimizing portfolios poses a complex challenge, and researchers have put forth various solutions and methods to address it. Machine learning models have played a significant role in the efforts of researchers to forecast future stock prices (Mehtab & Sen, 2021; Mehtab & Sen, 2020a; Mehtab & Sen, 2019; Mehtab et al., 2021; Sen, 2018a; Sen & Datta Chaudhuri, 2017a). The utilization of deep learning architectures and algorithms has led to enhancements in the predictive accuracy of these models (Sen & Mehtab, 2021b; Mehtab & Sen, 2021; Mehtab & Sen, 2020a; Mehtab & Sen, 2020b; Mehtab & Sen, 2019; Mehtab et al., 2021; Mehtab, et al., 2020; Sen, 2018a; Sen & Mehtab, 2021a; Sen & Mehtab, 2021b; Sen et al., 2021a; Sen et al., 2021b; Sen et al., 2021i; Sen et al., 2020; Sen & Mehtab, 2022b; Mehtab & Sen, 2019). In the realm of stock price prediction and portfolio design, time series decomposition-based statistical and econometric methods enjoy significant popularity as well (Sen, 2022a; Sen, 2018b; Sen, 2017; Sen & Datta Chaudhuri, 2018; Sen & Datta Chaudhuri, 2017b; Sen & Datta Chaudhuri, 2016a; Sen & Datta Chaudhuri, 2016b; Sen & Datta Chaudhuri, 2016c; Sen & Datta Chaudhuri, 2016d; Sen & Datta Chaudhuri, 2015).

The classical mean-variance optimization method stands out as the most widely recognized approach for portfolio optimization (Sen & Mehtab, 2022a; Sen et al., 2021e; Sen et al., 2021g; Sen et al., 2021h; Sen & Sen, 2023). Numerous researchers have put forward alternative methods for portfolio optimization, diverging from the traditional mean-variance approach. Prominent among these alternatives are eigen portfolios, which involve employing principal component analysis (Sen & Dutta, 2022b; Sen & Mehtab, 2022a), risk parity-based techniques (Sen & Dutta, 2022a; Sen & Dutta, 2022c; Sen & Dutta, 2021; Sen et al., 2021c; Sen et al., 2021f), and swarm intelligence-based approaches (Thakkar & Chaudhuri, 2021).

The literature on portfolio rebalancing is also very rich. Numerous propositions have been made by researchers for rebalancing portfolios to optimize their risk-adjusted returns. Some of such propositions are briefly discussed below.

Chaweewanchon & Chaysiri (2021) investigated the practical effectiveness of the traditional Markowitz portfolio optimization strategy with and without rebalancing was investigated by. The authors assessed the results in terms of the Sharpe ratio, portfolio return, and minimal risk. They also contrasted these findings with those rebalancing that involved transaction costs. The approach includes analyzing the 50-stock Stock Exchange of Thailand 50 Index (SET50)'s historical closing prices from January 2018 to December 2021. The outcomes demonstrated that a portfolio with a rebalancing strategy outperformed a portfolio without one.

Guo and Ryan (2021) used a rolling two-stage stochastic program to contrast time series momentum techniques with mean-risk optimization models. The authors divided investments between a market index and a risk-free asset to create future return possibilities based on a momentum-based stochastic process model. To generate trading signals using a modified momentum measure while adjusting the position of the risky asset to manage the conditional value-at-risk (CVaR) of return, a novel hybrid approach known as time series momentum strategy controlling downside risk (TSMDR) is developed. TSMDR outperforms conventional approaches. The findings showed that while time series momentum values and weighted moving averages both better reflect stock market trends, mean-risk strategies outperform risk parity techniques in terms of returns.

Darapaneni et al. (2020) designed a Q-Learning-based reinforcement learning framework that learns market patterns to trade in financial assets. The objective of the reinforcement learning agent was to maximize the fund value using portfolio returns net of transaction costs. The authors selected 15 Indian financial assets, including equity sectoral indices, government security indices, and gold spot prices, and trained the agent with the simple moving averages, 52-week stochastic indicators, and price change momentum indicators of their respective financial assets. It was found that most of the agents were successful in reducing the maximum drawdown and standard deviation.

Pai (2018) proposed an active portfolio rebalancing model to maximize the diversification ratio and the expected portfolio return. It considers non-linear constraints such as risk budgeting and other investor preferential constraints specified for the original portfolio, the transaction costs for rebalancing, and the rebalanced portfolio risk. The portfolio rebalancing model is a multi-objective non-convex non-linear constrained fractional programming problem, which is challenging to solve directly using traditional methods. To solve the multi-objective, non-convex, non-linear constrained non-integer programming optimization problem, the author used a multi-objective metaheuristics method.

Lejeune & Prasad (2017) presented a novel dynamic portfolio rebalancing technique that operates within the mean-risk framework. In this approach, the risk aversion coefficient is adjusted based on market trend information, which is computed using a technical indicator. The authors used Gini's mean difference as the risk measure and the moving average as the technical indicator. To validate their proposed scheme, the authors performed a comprehensive empirical evaluation using S&P 500 market data and a rolling horizon approach. The results demonstrated that the proposed time-varying risk-aversion adjustment-based portfolio rebalancing strategy yields higher returns compared to a strategy that employs a fixed risk-aversion coefficient.

Maree & Omlin (2022) developed an innovative utility function that combines the Sharpe ratio, which represents risk, with the environmental, social, and governance score (ESG), which represents sustainability. The authors argued that the multi-agent deep deterministic policy gradients (MADDPG) method fails to identify the optimal policy due to flat policy gradients. To solve the problem, the authors proposed a genetic algorithm for optimizing the parameters of the gradient descent. The results showed that the proposed algorithm outperforms MADDPG.

Strub (2017) argued that mixed-integer linear programming (MILP) approaches to portfolio rebalancing very often result in portfolios with negative excess returns or high tracking errors. Researchers have proposed several mixed-integer linear programming (MILP) formulations to address this problem. To address these problems, the author proposed a novel MILP scheme that carries out the rebalancing task by replicating a carefully constructed tracking target over a historical in-sample period. The experimental results demonstrated that the portfolios designed using the proposed approach achieve high excess returns and low tracking errors.

Albertazzi et al. (2021) analyzed cross-sectional heterogeneity in how the financial portfolios of different sectors of the European economy were affected by the purchase program. The European Central Bank's large-scale asset purchase program, while primarily targeting safe assets, also aimed to influence the prices of risky assets. The study found ample evidence of portfolio rebalancing with countries that were more vulnerable to macroeconomic imbalances and relatively high-risk premia, exhibited an increasing tendency to shift towards riskier securities. On the other hand, for less vulnerable countries, a rebalancing trend towards bank loans was observed.

Horn & Oehler (2020) examined whether households would benefit from an automated rebalancing service that includes frequently tradable assets such as real estate funds, articles of great value, and cash(-equivalents) in addition to stocks and bonds. The authors analyzed real-world household portfolios derived from the German Central Bank's Panel on Household Finances (PHF)-Survey. The work involved the computation of the increase/decrease in portfolio performance that households would have achieved by employing rebalancing strategies instead of a buy-and-hold strategy between September 2010 and July 2015. It also investigates whether certain sociodemographic and socioeconomic characteristics of households would have influenced the benefits of portfolio rebalancing. The results indicated no significant positive impact of the automated rebalancing approach as no subgroup of households was found to have significantly outperformed another for active rebalancing.

Hilliard & Hilliard (2018) examined the returns from rebalanced and buy-and-hold portfolios that consist of the same stocks. The authors derived theoretical properties using Jensen's Inequality and Hölder's Defect Formula. It was observed that the rebalancing portfolios reduce total return volatility, while buy-and-hold strategies yield higher expected returns. In general, the results indicated that while rebalancing reduces volatility and momentum effects, the buy-and-hold strategy outperformed them due to the relatively higher returns offered by stocks compared to the risk-free asset.

Cuthbertson et al. (2016) studied the effects of portfolio rebalancing on their returns and risks. The authors aimed to identify the misleading claims associated with rebalanced strategies and demonstrated, through theoretical analysis and simulations, that the apparent advantages of rebalanced strategies over infinite time horizons do not accurately reflect their performance over finite time horizons.

Guo & Ryan (2023) used a rolling two-stage stochastic program to compare mean-risk optimization models with time series momentum strategies to analyze the trade-off between risk and return in financial investments. By backtesting the allocation of investment between a market index and a risk-free asset, the authors generated future return scenarios based on a momentum-based stochastic process model. The proposed scheme, known as the time series momentum strategy controlling downside risk (TSMDR), was found to outperform traditional approaches by generating trading signals using a modified momentum measure while adjusting the risky asset position to control the conditional value-at-risk (CVaR) of return.

Hagiwara & Harada (2017) argued there is a need for recombining assets and changing the proportion of asset allocation in a portfolio through rebalancing as the performance of a portfolio may not be sustainable over a long period. The authors proposed a dynamic rebalancing scheme for portfolios that works on instance-based policy optimization, based on the changes in market conditions.

Jigang & Chang (2020) proposed a portfolio rebalance framework that integrates machine learning models into the mean-risk portfolios in multi-period settings with risk-aversion adjustment. In each period, the risk-aversion coefficient is adjusted automatically according to market trend movements predicted by machine learning models. The results showed that the XGBoost model is the most accurate one in predicting market movements. On the other hand, the proposed rebalance strategy was found to generate portfolios with superior out-of-sample performances.

Fischer et al. (2021) investigated the dynamics of international portfolio equity flows and their time variation. The authors extended the empirical model of Hau and Rey (2004) by incorporating a Markov regime-switching scheme into the structural vector autoregression (VAR) model. The model is estimated using monthly data from 1995 to 2018, focusing on equity returns, exchange rate returns, and equity flows between the United States and advanced and emerging market economies. The findings suggest that a two-state model is favored by the data, where coefficients and shock volatilities switch jointly.

Laher et al. (2021) proposed a deep learning-based portfolio management model for forecasting weekly returns of financial time series. The model, built on the principle of late fusion of an ensemble of forecast models, is a modified version of the standard mean-variance optimizer that has the capability of handling transaction costs in multi-period trading. The empirical results demonstrate that the portfolio management tool outperforms the equally weighted portfolio benchmark and the buy-and-hold strategy, utilizing both Long Short-Term Memory (LSTM) and Gated Recurrent Unit (GRU) forecasts.

Delpini et al. (2020), studied a real-world holdings network and compared it with various alternative scenarios involving randomization and rebalancing of the original investments. The scenarios were generated using algorithms that adhere to the global constraints imposed by the number of outstanding shares in the market. The authors examined both fixed-diversification models and diversification-maximizing replicas. The results indicated that real portfolios tend to be poorly diversified, while there is a correlation between portfolio similarity and systemic fragility. It was also demonstrated that rebalancing often leads to significant diversification gains, but it also renders the network more vulnerable to unselective shocks.

Bernoussi & Rockinger (2023) argued that when transaction costs are absent and returns are independent, a buy-and-hold strategy is expected to generate higher returns than a fixed-weight strategy. The fixed-weight strategy involves regularly readjusting or rebalancing the portfolio weights to an initial level. However, the buy-and-hold strategy's higher expected return is accompanied by increased volatility. Consequently, the ranking of the Sharpe ratio varies depending on the statistical moments of the assets. The authors explored the concept of Maximum Drawdown and discussed factors that influence the ranking of the Sharpe ratio. Furthermore, the authors also analyzed several realistic portfolios encompassing risk-free assets, bonds, stock indices, commodities, and real estate, and found that rebalanced portfolios yield higher returns in the majority of the cases.

Tunc et al. (2013) designed optimal investment strategies in a stock market with a limited number of assets from a signal processing perspective. The authors proposed a portfolio selection algorithm for maximizing the expected cumulative wealth in discrete-time markets with two assets. The approach utilizes the concept of 'threshold rebalanced portfolios', that only trigger trades when certain thresholds are crossed.

Kim & Lee (2020) investigated the portfolio choices of equity mutual funds in emerging markets with varying degrees of financial market integration. The authors analyzed the monthly holdings of 385 mutual funds from 1999 to 2017 and observed that these funds typically employ portfolio rebalancing strategies in response to changes in equity returns. Furthermore, the study revealed that the inclination to rebalance is higher in stock markets that exhibit greater financial integration with the global market. The presence of high market liquidity and low regulatory barriers, which are indicative of financial integration, emerge as significant factors driving active rebalancing in emerging markets.

The current work presents an adaptable rebalancing approach for stock portfolios. While the approach can be adapted to either a daily, monthly, or yearly basis, the performance of the rebalancing approach has been studied for yearly rebalancing on stocks chosen from ten important sectors of the Indian stock market. To the best of the knowledge and belief of the authors, no such studies have been done so far in this direction. Hence, the results of this work are expected to be useful to financial analysts and investors interested in the Indian stock market.

## Some Theoretical Background

In this section, some background theories are discussed that will be needed for a proper understanding of the methodology used in the work and the subsequent analysis of the results. In the following, some important terms used in portfolio analysis are defined and their usefulness is explained.

**Annual return:** The annual return is the gain that an investment yields during a given timeframe expressed as an annual percentage that considers the effects of time. This annualized rate of return is assessed with the initial investment amount and is represented as a geometric mean rather than a simple arithmetic average. In other words, an annual return gives the average yearly growth of an investment over a specified period. When assessing an investment's performance over an extended period or comparing two investments, an annual return can offer more valuable insights than a simple return. The compound annual growth rate (CAGR) of an investment is given by (1)

$$CAGR = \left( (\frac{Final\ value}{Initial\ amount})^{\frac{1}{No\ of\ years}} \right) - 1 \qquad (1)$$

**Cumulative return:** The cumulative return of an investment represents the total amount of gain or loss that the investment has experienced over time, regardless of the period involved. This cumulative return (CR) is typically expressed as a percentage and is derived from (2)

$$CR = \frac{CPI - OPI}{OPI} \qquad (2)$$

In (2), CPI refers to the *current price of investment*, and OPI stands for the *original price of investment*. The cumulative return signifies the overall change in the investment's price over a specified period, reflecting a combined return, rather than an annualized one. As an illustration, if an investor invested an amount of $10,000 in the stock of ABC Inc. and, after 10 years, the value of the stocks grew to $48,000, this would represent a cumulative return of 380%. This calculation is based solely on the initial and final investment values, without factoring in taxes or reinvested dividends.

**Annual volatility:** Annualized volatility is a statistical metric that gauges the spread or variability in the returns of a financial instrument during a specific time frame, presented as an annualized standard deviation. Its primary purpose is to quantify the level of risk associated with an investment or portfolio by indicating the expected degree of fluctuation in the investment's value over a set period. Higher annualized volatility values signify greater investment risk. Typically, this measure is computed using historical return data and is expressed as a percentage. Investors frequently rely on annualized volatility to inform their investment decisions.

*Maximum drawdown*: A maximum drawdown (MDD) represents the most substantial observed decline in the value of a portfolio, measured from its highest point to its lowest point before it eventually reaches a new peak. Maximum drawdown serves as an essential indicator of the potential downside risk associated with a portfolio over a specified period. It helps investors assess how much loss their investments might experience at their worst moments before recovering to higher levels.

It focuses on identifying the most significant decline from a peak to a trough within a portfolio, emphasizing the magnitude of the largest loss without regard to the frequency of such occurrences. It's important to recognize that while MDD provides valuable insights into the depth of potential losses, it does not provide information regarding the duration it takes for an investor to recover from those losses or whether the investment eventually rebounds to its previous levels. Therefore, MDD is just one component of a more comprehensive risk assessment, and investors should consider additional factors when evaluating the overall risk and performance of an investment or portfolio.

Maximum drawdown (MDD) is indeed a valuable indicator for assessing the relative riskiness of different stock screening strategies. It is especially relevant because it places a strong emphasis on capital preservation, which is a primary concern for most investors. Even if two screening strategies have similar average outperformance, tracking error, and volatility, their maximum drawdowns concerning the benchmark can vary significantly.

A low maximum drawdown is generally preferred because it signifies that losses incurred from the investment were limited. In an ideal scenario where an investment never experienced any losses, the maximum drawdown would be zero, indicating the preservation of capital. Conversely, the worst possible maximum drawdown would be -100%, implying that the investment has become entirely worthless, which is a situation investors typically want to avoid. Therefore, assessing and comparing maximum drawdowns can be a crucial aspect of making informed investment decisions, particularly for those seeking to manage risk effectively.

*Sharpe ratio*: The Sharpe ratio is a metric that evaluates the relationship between an investment's return and its level of risk. It was developed by economist William F. Sharpe in 1966, building upon his research on the Capital Asset Pricing Model (CAPM). Initially, Sharpe referred to this ratio as the "reward-to-variability ratio."

The Sharpe ratio serves as a tool for assessing the performance of an investment by considering not only the return it generates but also the amount of risk or volatility associated with that return. In essence, it helps investors gauge whether the potential reward of an investment justifies the level of risk taken to achieve that return. A higher Sharpe ratio typically suggests a more favorable risk-return trade-off, making it a valuable measure for evaluating and comparing different investment options.

The numerator of the Sharpe ratio represents the difference between the realized (or expected) returns of an investment and a chosen benchmark. The benchmark is typically the risk-free rate of return or the performance of a specific investment category. This difference measures the excess return generated by the investment above the risk-free rate or the benchmark. The denominator is the standard deviation of the investment's returns over the same period. Standard deviation is a statistical measure of the dispersion or volatility of those returns. It quantifies the investment's risk, with higher standard deviations indicating greater volatility and hence, higher risk.

By dividing the excess return (numerator) by the risk or volatility (denominator), the Sharpe ratio quantifies how much additional return an investment generates for each unit of risk taken. In other words, it assesses the efficiency of an investment in generating returns relative to the risk involved. A higher Sharpe ratio is generally considered more favorable because it implies better risk-adjusted performance.

The top part of the Sharpe ratio, which represents the total return difference compared to a benchmark ($R_p$ - $R_f$), is computed as the mean of the return differences observed in each of the smaller time intervals that collectively constitute the total period. For instance, when calculating the numerator for a 10-year Sharpe ratio, you would take the average of the 120 monthly return differences between a fund and an industry benchmark.

**Calmar ratio:** The Calmar ratio serves as an indicator of the performance of investment funds like hedge funds and commodity trading advisors (CTAs). It calculates the fund's average compounded annual rate of return relative to its maximum drawdown. A higher Calmar ratio indicates superior performance in terms of risk-adjusted returns, typically evaluated over 36 months. This ratio was first introduced in 1991 by Terry W. Young, a fund manager based in California.

One of the strengths of the Calmar ratio is its reliance on the maximum drawdown as a risk measure. It is more easily comprehensible compared to other more abstract risk metrics, making it a preferred choice for some investors. Additionally, its standard three-year evaluation period makes it more reliable compared to shorter-term metrics that might be influenced by short-term market fluctuations.

However, the Calmar ratio's focus on drawdown means it has a limited perspective on risk compared to other metrics, and it does not consider general market volatility. This limitation reduces its statistical significance and overall utility. Nevertheless, the Calmar ratio, with its emphasis on risk-adjusted returns, is one of many potential metrics for assessing investment performance, albeit one of the lesser-known measures of risk-adjusted returns.

**Sortino ratio:** The Sortino ratio is a variation of the Sharpe ratio that differentiates harmful volatility from total overall volatility by using the asset's standard deviation of negative portfolio returns—downside deviation—instead of the total standard deviation of portfolio returns. The Sortino ratio takes an asset or portfolio's return subtracts the risk-free rate, and then divides that amount by the asset's downside deviation. The ratio was named after Frank A. Sortino.

The Sortino ratio is a useful way for investors, analysts, and portfolio managers to evaluate an investment's return for a given level of bad risk. Since this ratio uses only the downside deviation as its risk measure, it addresses the problem of using total risk, or standard deviation, which is important because upside volatility is beneficial to investors and is not a factor most investors worry about.

Just like the Sharpe ratio, a higher Sortino ratio result is better. When looking at two similar investments, a rational investor would prefer the one with the higher Sortino ratio because it means that the investment is earning more return per unit of the bad risk that it takes on.

***Omega ratio***: The Omega Ratio serves as a performance assessment tool utilized in the realm of finance and investment to gauge the trade-off between risk and return in each investment or portfolio. This metric evaluates the probability of attaining a target return in comparison to the potential downside risk. A higher Omega Ratio indicates a more advantageous risk-return profile, suggesting a more favorable investment.

The Omega Ratio was introduced in the early 2000s by Con Keating and William Shadwick as an alternative to conventional risk measurements like the Sharpe Ratio and the Sortino Ratio. It was specifically developed to overcome the limitations of these measures, particularly their reliance on assumptions of normality in return distributions. Portfolio managers, financial advisors, and individual investors widely employ the Omega Ratio to assess the balance between risk and reward across various investment choices. It aids in making more informed decisions and contributes to the overall optimization of investment portfolios.

The threshold return, also known as the minimum acceptable return, is a predetermined level of return that an investor aims to attain. It serves as a reference point for assessing the performance of an investment or portfolio.

This ratio offers a holistic assessment of risk and reward, considering an investor's particular target return. It allows for a customized evaluation of an investment's performance relative to the investor's specific objectives.

An Omega Ratio greater than 1 indicates that the investment has a higher probability of achieving the target return than experiencing a loss. Conversely, an Omega Ratio of less than 1 suggests that the investment is more likely to underperform the target return. Higher

Omega Ratios are generally preferable, as they represent better risk-adjusted performance.

*Tail ratio*: It is the ratio of the absolute value of the ratio of the right (95%) tail to the left tail (5%) of the distribution of the daily return. For computing this ratio, we select the 95th and 5th quantiles for the distribution of daily return and then divide to obtain the absolute value. The ratio signifies how many times the return earned is greater than the loss. A value of greater than 1 for the tail ratio is desirable for a portfolio.

*Skewness*: Skewness is a statistical measure that characterizes the asymmetry or lack of symmetry in the distribution of a set of data points. In the context of a portfolio's return, skewness assesses the shape of the distribution of those returns and how much the distribution is deviant from a normal distribution.

In context to the returns of a portfolio, skewness has the following significance.

*Positive Skewness*: If the portfolio's return distribution is positively skewed, it means that the distribution is skewed to the right. In this case, most returns tend to be clustered on the left side of the distribution (below the mean), with relatively few extreme positive returns on the right side (above the mean). Positive skewness suggests that while the portfolio generally has modest gains, it occasionally experiences large gains.

*Negative Skewness*: Conversely, if the portfolio's return distribution is negatively skewed, it indicates a leftward skew. In this scenario, most returns cluster on the right side of the distribution (above the mean), with relatively few extreme negative returns on the left side (below the mean). Negative skewness suggests that the portfolio generally has modest losses, but it occasionally incurs large losses.

*Zero Skewness*: A skewness of zero implies that the distribution of returns is symmetric. In this case, the returns are evenly distributed around the mean without any pronounced skew to one side or the other.

Understanding the skewness of a portfolio's return distribution is crucial for risk assessment and portfolio management. It provides insights into the likelihood of extreme returns, both positive and negative, which can help investors and portfolio managers make informed decisions about risk tolerance, hedging strategies, and asset allocation. Positive skewness may be desirable for certain investment strategies, while negative skewness may indicate higher risk and potential for substantial losses.

**Kurtosis:** Kurtosis is a statistical measure that describes the distribution of a dataset, particularly focusing on the tails of the distribution. In the context of a portfolio's returns, kurtosis can help you understand the shape of the distribution and the likelihood of extreme returns (both positive and negative). There are two main types of kurtosis other than the kurtosis exhibited by normally distributed data. Normally distributed data is said to be mesokurtic. Non-normal data exhibit either leptokurtic or platykurtic behavior.

*Leptokurtic*: A distribution with positive kurtosis is referred to as leptokurtic. This means that the distribution has fatter (i.e., heavier) tails and a sharper peak than a normal distribution. In the context of a portfolio's returns, a leptokurtic distribution suggests that there is a higher probability of extreme returns (both positive and negative) compared to a normal distribution. This indicates that the portfolio's returns may be more volatile and have a higher risk of outliers. A leptokurtic distribution has a kurtosis greater than 3.

*Platykurtic*: A distribution with negative kurtosis is referred to as platykurtic. This means that the distribution has thinner (i.e., lighter) tails and a flatter peak than a normal distribution. In the context of a portfolio's returns, a platykurtic distribution suggests that there is a lower probability of extreme returns compared to a normal distribution. This indicates that the portfolio's returns may be less volatile and have a lower risk of outliers. A platykurtic distribution has a kurtosis value of less than 3.

A mesokurtic distribution (i.e., normally distributed data) has a kurtosis value of 3.

Kurtosis is just one measure of a portfolio's risk and should be considered alongside other risk metrics and analysis techniques to get a comprehensive understanding of its return distribution. High kurtosis may indicate a higher risk of extreme events, but it should be evaluated in conjunction with other factors such as skewness, and volatility.

**Stability:** It determines the *r-squared* value of a linear fit to the cumulative log returns of a portfolio. This is not a standard term in the portfolio literature. The *pyfolio* library which has been used in this work uses the term "stability" to refer to the *r-squared* value of a linear regression model fitted into the cumulative log return values to time (Kutner et al., 2004). The *r-squared* value, also known as the coefficient of determination, is a statistical measure that indicates the proportion of the variance in the dependent variable (in this case, the cumulative log return) that can be explained by the independent variable (in this case, time) in a regression analysis.

**Value at risk:** Value at Risk (VaR) is a numerical measure that quantifies the potential financial losses that can occur within a company, investment portfolio, or position over a defined period. This statistic finds its primary application in the financial industry, particularly among investment and commercial banks, where it is used to assess the magnitude and likelihood of prospective losses in their institutional portfolios. Risk managers employ VaR as a tool to gauge and manage the degree of risk exposure. VaR calculations can be applied to individual positions or entire portfolios, and they can also be used to assess the overall risk exposure at the firm level. This flexibility allows risk managers to tailor their risk assessment and control efforts to specific needs, whether at the asset level, portfolio level, or across the entire organization.

VaR modeling aims to evaluate the potential for loss within the entity under scrutiny and the likelihood of that defined loss occurring. To measure VaR, one considers the possible amount of loss, the probability of that loss occurring, and the time frame involved.

For instance, let us say a financial firm determines that a particular asset has a one-month VaR of 2% with a 3% probability, indicating a 3% chance that the asset's value will decrease by 2% during the one month. Converting this 3% chance into a daily ratio suggests that there is a one-day-per-month likelihood of a 2% loss.

By conducting a firm-wide VaR assessment, institutions can assess the cumulative risks stemming from combined positions held across various trading desks and departments within the organization. With the insights derived from VaR modeling, financial institutions can determine if they have sufficient capital reserves to cover potential losses or if the presence of higher-than-acceptable risks necessitates the reduction of concentrated holdings.

**Alpha:** *Alpha* (α) is a concept in the realm of investing that characterizes an investment strategy's capacity to outperform the market or its competitive advantage. It is frequently described as the "excess return" or the "abnormal rate of return" relative to a benchmark, once the influence of risk is taken into account. *Alpha* is commonly employed alongside beta (represented by the Greek letter β), which assesses the overall volatility or risk associated with the broader market, often referred to as systematic market risk.

*Alpha* serves as a critical metric in finance for assessing performance, specifically to determine whether a strategy, trader, or portfolio manager has achieved returns that surpass the market or another designated benchmark during a specific period. It essentially quantifies the active return on investment, measuring how well an investment has performed relative to a market index or a benchmark that is considered representative of the overall market's performance.

*Alpha* represents the excess return of an investment when compared to the return of a benchmark index. This measurement can either be positive or negative and is a result of active investing, reflecting the skill or strategy employed by an investor. In contrast, *beta* is a measure that can be obtained through passive index investing and typically represents the overall market risk.

Active portfolio managers aim to generate alpha within diversified portfolios, utilizing diversification to minimize unsystematic risk. *Alpha*, in this context, signifies the performance of a portfolio concerning a benchmark, and it is commonly viewed as the value that a portfolio manager contributes to or detracts from a fund's overall return.

In simpler terms, *alpha* represents the investment return that is independent of the broader market's movements. Thus, an *alpha* of zero would suggest that the portfolio or fund is closely mirroring the benchmark index, and the manager has neither added nor subtracted any extra value in comparison to the broader market.

**Beta:** Beta (β) is a metric that quantifies the level of volatility or systematic risk associated with a particular security or portfolio in comparison to the broader market, typically represented by a benchmark index like the S&P 500 on a global scale or the NIFTY50 index in the context of India. Stocks with beta values exceeding 1.0 are generally considered to be more volatile than the benchmark, signifying a higher degree of price fluctuations relative to the overall market.

A *beta* coefficient serves as a measure that assesses the level of volatility in an individual stock with the systematic risk of the overall market. In statistical terms, *beta* corresponds to the slope of a line derived from a regression analysis of data points. In the context of finance, each of these data points represents the returns of an individual stock plotted against the returns of the entire market.

*Beta* effectively characterizes how a security's returns behave in response to fluctuations in the market. To calculate a security's beta, you divide the product of the covariance between the security's returns and the market's returns by the variance of the market's returns over a specified period. This calculation quantifies the stock's sensitivity to market movements and provides insights into how it tends to perform in various market conditions.

# Methodology

This section presents the details of the data used and discusses the methodology followed in this work especially focusing on the steps involved in designing the portfolio rebalancing scheme. The methodology involves the following steps.

**(i) *Choice of the sectors for analysis:*** Ten important sectors are first selected from those listed in the NSE. The chosen ten sectors are (i) *auto*, (ii) *banking*, (iii) *consumer durables*, (iv) *fast-moving consumer goods* (FMCG), (v) *information technology* (IT), (vi) *metal*, (vii) *pharma*, (viii) *private banks*, (ix) *PSU banks*, and (xv) *realty*. The monthly reports of the NSE identify the ten stocks with the maximum free-float capitalization from each sector. In this work, the report published on January 4, 2021, is used for identifying the ten stocks from each of the ten sectors (NSE Website).

**(ii)** *Extraction of historical stock prices from the web:* From the Yahoo Finance website, the historical daily prices of the stocks are extracted from January 4, 2021, to September 20, 2023, using the *download* function of the *yfinance* module of Python. The portfolios are built on the records from July 1, 2019, to June 30, 2022. The historical values of the NIFTY 50 index are also downloaded for the same period as the benchmark index. The adjusted close prices of the stocks are used for forming the portfolios. Since the current work does not aim to optimize the portfolios following any specific method, an equal-weight allocation is made on the first day for each stock in a given portfolio. The historical prices from January 4, 2021, to June 30, 2022, are used as the in-sample data for training the portfolios, while the stock price records from July 1, 2022, to September 20, 2023, are used as the out-of-sample data for testing the performances of the portfolios.

**(iii)** *Design of equal-weight portfolios for the sectors:* For each sector, on the first day (i.e., on January 4, 2021) an equal-weight portfolio is built. For this, an amount of Indian Rupees (INR) 1,00,000 is allocated for each stock in the portfolio of a sector. In other words, the initial investment of INR 10,00, 000 is made for each of the ten sector-specific portfolios. Based on the prices of the stocks on January 4, 2021, the initial number of shares for each stock in each portfolio is computed using a Python function. Since the number of shares of a stock needs to be an integer, the results of the division of the initial investment amount by the prices of the stocks on January 4, 2021, are rounded off to the nearest integer values.

**(iv)** *Designing the rebalanced portfolios:* The rebalancing of the portfolios is done based on the output of a Python function that computes the log-returns of the portfolios at daily, monthly, and yearly intervals. As the prices of the constituting stocks of a portfolio vary the number of shares of the stocks need to be adapted based on the changes in their prices to maximize the return. Rebalancing daily is not a feasible option as the transaction costs associated with the rebalancing will be too high in comparison to the return yielded by the rebalanced portfolio. Hence, monthly and yearly rebalancing is usually done in the real world. The Python function is made adaptable by passing a variable parameter that is set to either "daily", monthly", or "yearly" based on the rebalancing type needed by the investor. In this chapter, the results are presented for portfolios of the ten sectors which are rebalanced yearly. The yearly-balanced portfolios of the ten sectors are compared for their performances on several metrics with the benchmark NIFTY 50 index.

**(v)** *Visual presentation of the performance of the portfolios:* To visualize the behavior of the portfolios and their performance, several graphs and charts are constructed using matplotlib and seaborn libraries of Python. The following graphs and charts are constructed for each portfolio: (a) the number of shares of each stock in the portfolio over the training (i.e., in-sample) and the test (i.e., out-of-sample) records, (b) the variation of the weights corresponding to each stock over the training and the test records, (c) the graphs of the cumulative daily return of the portfolio over the training and the test records and its comparison with the cumulative daily return based on the benchmark NIFTY 50 index, and (d) the plot of the statistical distribution of the monthly returns over the training, the bar plots of the mean annual returns over the training and the test records, and the box-plots of the daily, monthly, and annual returns over the entire period.

**(vi)** *Evaluation of the performance of the portfolios:* In the final step, using the *create_full_tear_sheet* function of the *pyfolio* library, several numerical metrics such as Sharpe ratio, Calmar ratio, Sortino ratio, Omega ratio, daily value at risk, alpha, and beta are computed to evaluate the performance of each portfolio. The performance of each portfolio is compared with the performance of the benchmark index of NIFTY 50. The values of the numerical metrics will help evaluate the sectors' performance and their comparative performance with the benchmark index of NIFTY 50.

## Experimental Results

This section presents the detailed results and analysis of the rebalancing strategy of the portfolios. The ten sectors which are studied in this work are the following (i) *auto*, (ii) *banking*, (iii) *consumer durables*, (iv) *FMCG*, (v) *IT*, (vi) *metal*, (vii) *pharma*, (viii)

*private banks*, (ix) *PSU banks*, and (x) *realty*. The rebalanced portfolios are implemented using Python 3.9.8 and its associated libraries *numpy*, *pandas*, *matplotlib*, *seaborn, yfinance, and pyfolio* are used in the implementation. The portfolio models are trained and tested on a computing system running on the Windows 11 operating system, with an Intel i7-9750H CPU, a clock speed of 2.60GHz, and 16 GB RAM.

In the following, the performances of the rebalanced portfolios of ten sectors are presented in detail. As a general observation, it is found that the in-sample records (January 2021 - June 2022) correspond to a bearish period in which the stock market t of India was in the initial phase of the recovery post-COVID-19 period. On the other hand, the out-of-sample records largely correspond to a bullish period in which the stock market had returned to its normal behavior. Hence, for all sectors, the performance of the rebalanced portfolios is superior on the out-of-sample records compared to the in-sample records.

For each sector, the portfolio was created on January 4, 2021based on an equal-weight allocation approach for its constituent ten stocks. The initial amount of investment for each stock was INR 1,00,000. Thus, the initial capital invested for each portfolio was INR 10,00,000.

*Auto sector:* As per the report published by the NSE on January 4, 2021, the ten stocks of the *auto* sector with the largest free-float market capitalization and their contributions (in percent) to the overall sectoral index are the following: (i) Maruti Suzuki India (MARUTI): 18.55, (ii) Mahindra & Mahindra (M&M): 18.30, (iii) Tata Motors (TATAMOTORS): 14.59, (iv) Bajaj Auto (BAJAJ-AUTO): 7.54, (v) Eicher Motors (EICHERMOT): 6.20, (vi) Hero MotoCorp (HEROMOTOCO): 5.22, (vii) TVS Motor Company (TVSMOTOR): 4.66, (viii) Tube Investment of India (TIINDIA): 4.10, (ix) Bharat Forge (BHARATFORG): 3.68, and (x) Ashok Leyland (ASHOKLEY): 3.35 (NSE Website). The ticker names of the stocks are mentioned in parentheses. The ticker name of a stock is its unique identifier for a given stock exchange.

Figure 2.1 shows how the number of shares for the stocks constituting the rebalanced *auto* sector portfolio varied over the entire period (i.e., including both in-sample and out-of-sample records). The initial number of shares for the stocks in the portfolio on January 4,

2021, were as follows: (i) MARUTI: 13, (ii) M&M: 142, (iii)TATAMOTORS: 537, (iv) BAJAJ-AUTO: 32, (v) EICHERMOT: 40, (vi) HEROMOTOCO: 35, (vii) TVSMOTOR: 207, (viii) TIINDIA: 126, (ix) BHARATFORGE: 189, and (x) ASHOKLEY: 1059.

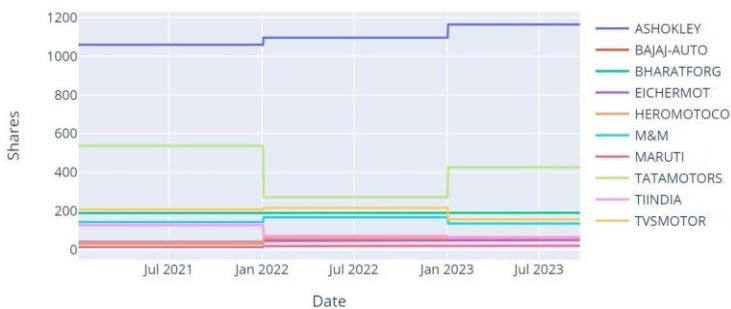

**Figure 2.1.** The daily number of shares of each stock in the *auto* sector portfolio from January 4, 2021, to September 20, 2023.

Figure 2.2 depicts how the weights corresponding to the stocks of the *auto* sector portfolio varied over the entire period. While the daily variation of weights is shown, the rebalancing was done only yearly.

Figure 2.3 exhibits the backtesting results of the portfolio performance on its cumulative returns and its comparison with the benchmark cumulative returns of the NIFTY 50 index. The rebalanced *auto* sector portfolio yielded a consistently higher cumulative return compared to the benchmark NIFTY 50 index.

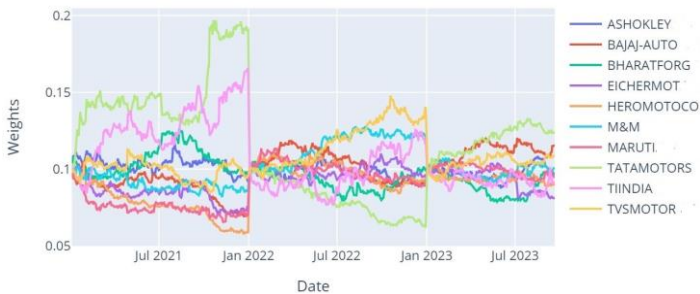

**Figure 2.2.** The daily allocation of weights to each stock of the *auto* sector portfolio from January 4, 2021, to September 20, 2023.

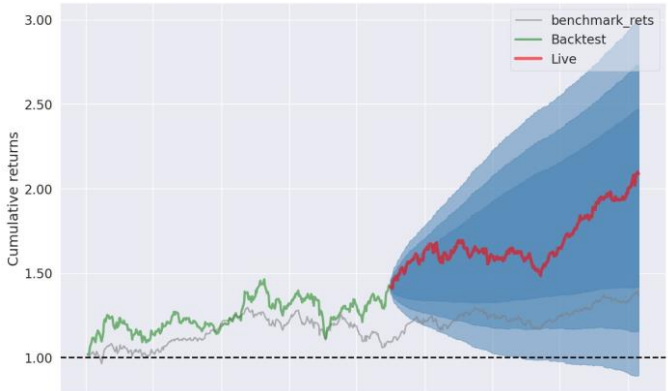

**Figure 2.3.** The cumulative return of the *auto* sector portfolio and the cumulative return of the benchmark index of NIFTY 50 from January 4, 2021, to September 20, 2023. The green, red, and gray lines indicate the cumulative returns for the in-sample records, out-of-sample records of the *auto* sector portfolio, and the benchmark NIFTY 50 index.

Figure 2.4 depicts several statistical features of the *auto* sector portfolio returns, including the monthly returns of the portfolios over the entire period, the annual returns, the distribution of monthly returns, and the box plots of the daily, monthly, and yearly returns.

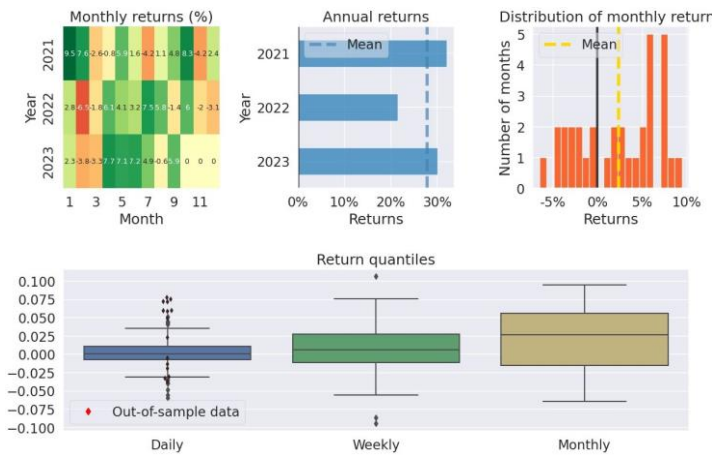

**Figure 2.4.** The statistical distribution and box plots of the daily, weekly, monthly, and annual returns of the rebalanced portfolio of the *auto* sector.

The detailed performance results of the rebalanced portfolio of the *auto* sector are presented in Table 2.1. It is observed that the portfolio yielded substantial annual and cumulative returns, while its annual volatility was moderate. The values of the Sharpe ratio, Sortino ratio, Calmar ratio, Omega ratio, and Tail ratio for both in-sample and out-of-sample are all greater than 1 indicating a good performance, particularly over the out-of-sample records (i.e., during the portfolio test period). A stability value of 0.87 indicates a good linear fit of the cumulative return with time. The skewness and the kurtosis values exhibit a negatively skewed and platykurtic behavior of the return. The daily value at risk indicates that with a probability of 0.95, the loss yielded by the portfolio did not exceed 2.95%, 1.76%, and 2.48%, for the in-sample records, out-of-sample records, and all records, respectively, over one day. The *alpha* values for the portfolio were positive for both in-sample and out-of-sample records indicating that the portfolio consistently outperformed the benchmark NIFTY 50 index. The portfolio yielded an excess return of 0.20% and 0.18% over the in-sample and out-of-sample records, respectively. The *beta* values of 1.03 for the in-sample records indicate the portfolio exhibited marginally higher volatility in comparison to the benchmark NIFTY 50 index. However, for the out-of-sample records and all records combined, the volatility was less as exhibited by the figures of 0.84 and 0.98, respectively.

**TABLE 2.1.** THE PERFORMANCE OF THE AUTO SECTOR PORTFOLIO ON THE IN-SAMPLE AND OUT-OF-SAMPLE DATA

| Metric | In-sample data | Out-of-sample data | Overall data |
|---|---|---|---|
| Annual return | 27.05% | 37.66% | 31.72% |
| Cumulative Return | 42.12% | 46.86% | 108.73% |
| Annual volatility | 24.28% | 15.04% | 20.63% |
| Max drawdown | -23.95% | -12.34% | -23.95% |
| Sharpe ratio | 1.11 | 2.20 | 1.44 |
| Calmar ratio | 1.13 | 3.05 | 1.32 |
| Sortino ratio | 1.63 | 3.46 | 2.15 |
| Omega ratio | 1.21 | 1.44 | 1.28 |
| Tail ratio | 1.05 | 1.21 | 0.98 |
| Skewness | -0.15 | -0.10 | -0.16 |
| Kurtosis | 1.40 | 1.08 | 2.16 |
| Stability | 0.48 | 0.65 | 0.87 |
| Daily value at risk | -2.95 | -1.76 | -2.48 |
| Alpha | 0.20 | 0.18 | 0.18 |
| Beta | 1.03 | 0.84 | 0.98 |

***Banking sector:*** As per the report published by the NSE on January 4, 2021, the ten stocks with the largest free-float market capitalization in the *banking* sector and their contributions (in percent) to the overall index of the sector are as follows: (i) HDFC Bank (HDFCBANK): 29.01, (ii) ICICI Bank (ICICIBANK): 23.14, (iii) Axis Bank (AXISBANK): 9.98, (iv) State Bank of India (SBIN): 9.83, (v) Kotak Mahindra Bank (KOTAKBANK): 9.61, (vi) IndusInd Bank (INDUSINDBK): 6.25, (vii) Bank of Baroda (BANKBARODA): 2.67, (viii) Federal Bank (FEDERALBNK): 2.32, (ix) AU Small Finance Bank (AUBANK): 2.30, and (x) IDFC First Bank (IDFCFIRSTB): 2.02 (NSE Website). The ticker names of the stocks are mentioned in parentheses.

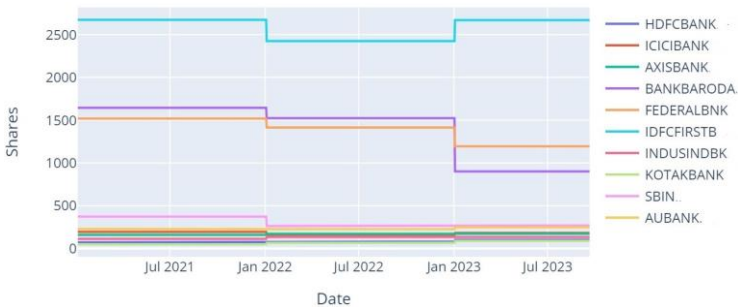

**Figure 2.5.** The daily number of shares of each stock in the *banking* sector portfolio from January 4, 2021, to September 20, 2023.

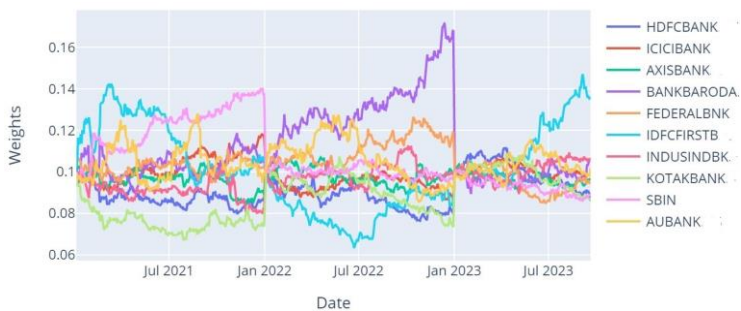

**Figure 2.6.** The daily allocation of weights to the stocks of the *banking* sector portfolio from January 4, 2021, to September 20, 2023.

Figure 2.5 shows how the number of shares for the stocks constituting the rebalanced *banking* sector portfolio varied over the entire period (i.e., including both in-sample and out-of-sample records). The initial number of shares for the stocks in the portfolio on January 4, 2021, were as follows: (i) HDFC Bank (HDFCBANK): 72, (ii) ICICI Bank (ICICIBANK): 194, (iii) Axis Bank (AXISBANK): 160, (iv) State Bank of India (SBIN): 373, (v) Kotak Mahindra Bank (KOTAKBANK): 50, (vi) IndusInd Bank (INDUSINDBK): 113, (vii) Bank of Baroda (BANKBARODA): 1644, (viii) Federal Bank (FEDERALBNK): 1519, (ix) AU Small Finance Bank (AUBANK): 228, and (x) IDFC First Bank (IDFCFIRSTB): 2673.

Figure 2.6 depicts how the weights corresponding to the stocks of the *banking* sector portfolio varied over the entire period. While the daily variation of weights is shown, the rebalancing was done only yearly.

Figure 2.7 exhibits the backtesting results of the portfolio performance on its cumulative returns and its comparison with the benchmark cumulative returns of the NIFTY 50 index. The rebalanced *banking* sector portfolio yielded a higher cumulative return compared to the benchmark NIFTY 50 index most of the time.

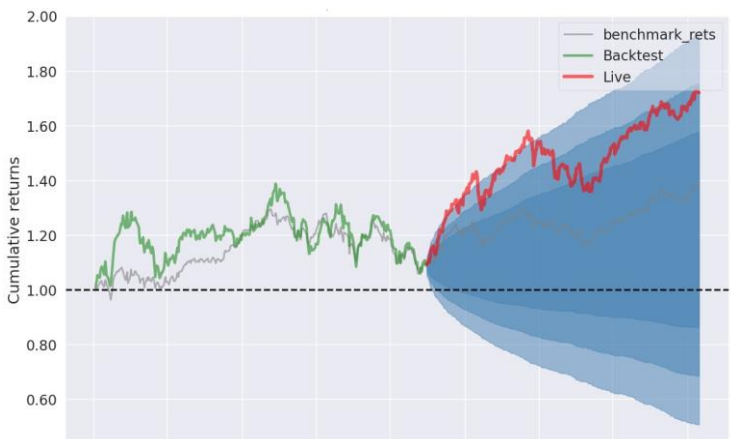

**Figure 2.7.** The cumulative return of the *banking* sector portfolio and the cumulative return of the benchmark index of NIFTY 50 from January 4, 2021, to September 20, 2023. The green, red, and gray lines indicate the cumulative returns for the in-sample records, out-of-sample records of the *banking* sector portfolio, and the benchmark NIFTY 50 index.

Figure 2.8 depicts several statistical features of the *banking* sector portfolio returns, including the monthly returns of the portfolios over the entire period, the annual returns, the distribution of monthly returns, and the box plots of the daily, monthly, and yearly returns.

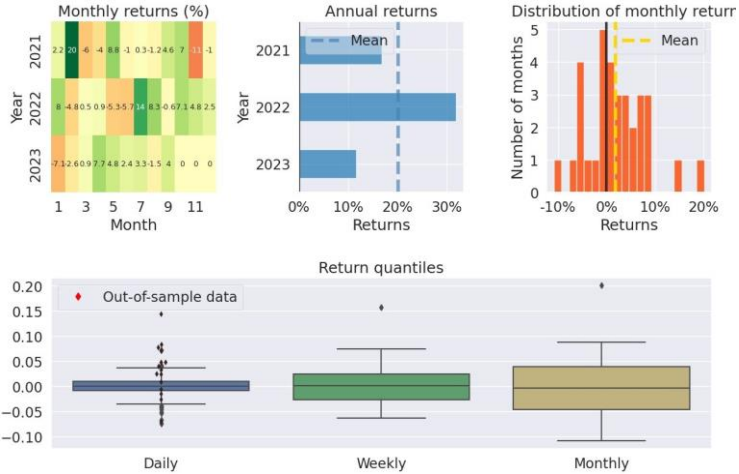

**Figure 2.8.** The statistical distribution and box plots of the daily, weekly, monthly, and annual returns of the rebalanced portfolio of the *banking* sector.

The detailed performance results of the rebalanced portfolio of the *banking* sector are presented in Table 2.2. It is observed that the portfolio yielded good annual and cumulative returns, especially on the out-of-sample records, while its annual volatility was moderate. The values of the Sharpe ratio, Sortino ratio, Calmar ratio, Omega ratio, and Tail ratio for the out-of-sample are all greater than 1 indicating a good performance over the portfolio test period. However, the ratios are not good for the in-sample records, indicating a bad performance of the portfolio over that period. A stability value of 0.70 indicates a good linear fit of the cumulative return with time. The skewness and the kurtosis values exhibit a negatively skewed and platykurtic behavior of the return. The daily value at risk indicates that with a probability of 0.95, the loss yielded by the portfolio did not exceed 3.28%, 1.99%, and 2.76%, for the in-sample records, out-of-sample records, and all records, respectively, over one day. The *alpha* values for the portfolio were positive for both out-of-sample records indicating that the portfolio outperformed the benchmark NIFTY 50 index on the test data. However, a negative value of *alpha* on the in-sample records indicates that the portfolio yielded a lower return than the benchmark index during that period. The *beta* values indicate the portfolio exhibited higher volatility in comparison to the benchmark NIFTY 50 index both on the in-sample and out-of-sample records.

**TABLE 2.2.** THE PERFORMANCE OF THE BANKING SECTOR
PORTFOLIO ON THE IN-SAMPLE AND OUT-OF-SAMPLE DATA

| Metric | In-sample data | Out-of-sample data | Overall data |
|---|---|---|---|
| Annual return | 5.89% | 46.36% | 22.50% |
| Cumulative Return | 8.76% | 58.09% | 71.94% |
| Annual volatility | 26.30% | 17.02% | 22.60% |
| Max drawdown | -23.68% | -13.96% | -23.68% |
| Sharpe ratio | 0.35 | 2.32 | 1.01 |
| Calmar ratio | 0.25 | 3.32 | 0.95 |
| Sortino ratio | 0.48 | 3.59 | 1.43 |
| Omega ratio | 1.06 | 1.49 | 1.19 |
| Tail ratio | 0.91 | 1.21 | 1.00 |
| Skewness | -0.42 | -0.19 | -0.45 |
| Kurtosis | 2.60 | 1.35 | 3.31 |
| Stability | 0.03 | 0.73 | 0.70 |
| Daily value at risk | -3.28% | -1.99% | -2.76% |
| Alpha | -0.01 | 0.19 | 0.07 |
| Beta | 1.23 | 1.15 | 1.21 |

*Consumer Durables sector:* As NSE's report published on January 4, 2021, the ten stocks from the *consumer durables* sector that have the largest free-float market capitalization and their contributions (in percent) to the overall index of the sector are as follows: (i) Titan Company (TITAN): 32.12, (ii) Havells India (HAVELLS): 14.80, (iii) Crompton Greaves Consumer Electricals (CROMPTON): 8.43, (iv) Voltas (VOLTAS): 8.38, (v) Dixon Technologies (DIXON): 8.30, (vi) Kajaria Ceramics (KAJARIACER): 4.65, (vii) Bata India (BATAINDIA): 4.39, (viii) Blue Star (BLUESTARCO): 4.23, (ix) Rajesh Exports (RAJESHEXPO): 2.90, and (x) Relaxo Footwears (RELAXO): 2.76 (NSE Website). The ticker names of the stocks are mentioned in parentheses.

Figure 2.9 shows how the number of shares for the stocks constituting the rebalanced *consumer durables* sector portfolio varied over the entire period (i.e., including both in-sample and out-of-sample records). The initial number of shares for the stocks in the portfolio on January 4, 2021, were as follows: (i) Titan Company (TITAN): 64, (ii) Havells India (HAVELLS): 111, (iii) Crompton Greaves Consumer Electricals (CROMPTON): 275, (iv) Voltas (VOLTAS): 123, (v) Dixon Technologies (DIXON): 36, (vi) Kajaria Ceramics (KAJARIACER): 145, (vii) Bata India (BATAINDIA): 66, (viii) Blue Star (BLUESTARCO): 257, (ix) Rajesh Exports (RAJESHEXPO): 206, and (x) Relaxo Footwears (RELAXO): 121.

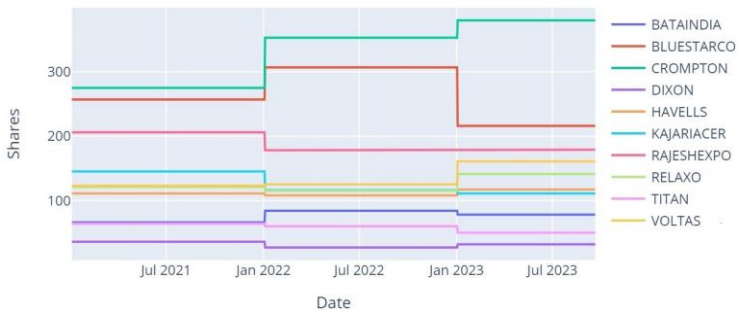

**Figure 2.9**. The daily number of shares of each stock in the *consumer durables* sector portfolio from January 4, 2021, to September 20, 2023.

Figure 2.10 depicts how the weights corresponding to the stocks of the *consumer durables* sector portfolio varied over the entire period. While the daily variation of weights is shown, the rebalancing was done only yearly.

Figure 2.11 exhibits the backtesting results of the portfolio performance on its cumulative returns and its comparison with the benchmark cumulative returns of the NIFTY 50 index. The rebalanced *consumer durables* sector portfolio yielded a higher cumulative return compared to the benchmark NIFTY 50 index most of the time.

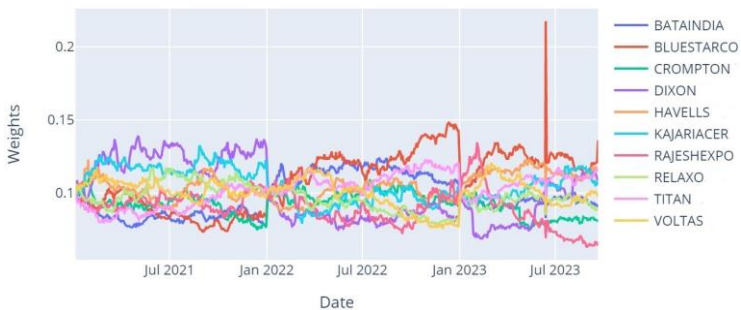

**Figure 2.10**. The daily allocation of weights to the stocks of the *consumer durables* sector portfolio from January 4, 2021, to September 20, 2023.

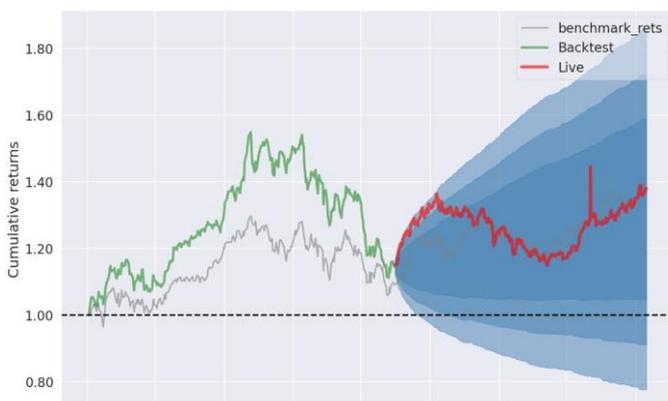

**Figure 2.11.** The cumulative return of the *consumer durables* sector portfolio and the cumulative return of the benchmark index of NIFTY 50 from January 4, 2021, to September 20, 2023. The green, red, and gray lines indicate the cumulative returns for the in-sample records, out-of-sample records of the *consumer durables* sector portfolio, and the benchmark NIFTY 50 index.

Figure 2.12 depicts several statistical features of the *consumer durables* sector portfolio returns, including the monthly returns of the portfolios over the entire period, the annual returns, the distribution of monthly returns, and the box plots of the daily, monthly, and yearly returns.

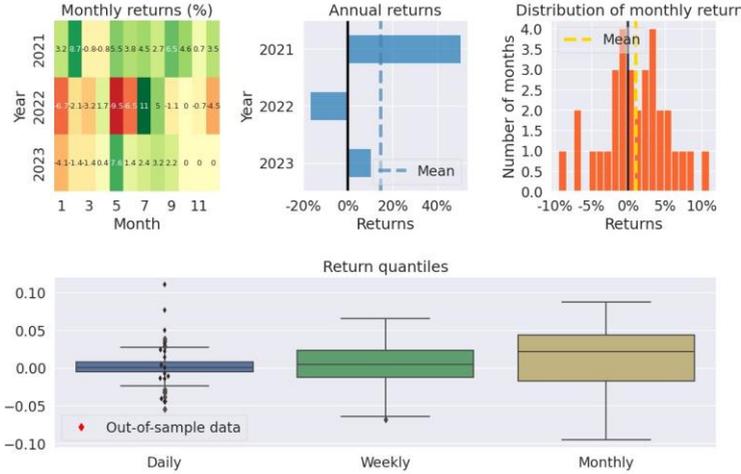

**Figure 2.12.** The statistical distribution and box plots of the daily, weekly, monthly, and annual returns of the rebalanced portfolio of the *consumer durables* sector.

The detailed performance results of the rebalanced portfolio of the *consumer durables* sector are presented in Table 2.3. It is observed that the portfolio yielded substantial annual and cumulative returns, while its annual volatility was moderate. The values of the Sharpe ratio are less than 1 for the in-sample and out-of-sample data indicating that the risk-adjusted returns are low. The Calmar ratio and Sortino ratio are also less than 1 for the in-sample data indicating an inadequate return in comparison to the risk. However, for the out-of-sample records, except for the Sharpe ratio, all other ratios yielded values that are greater than 1 indicating superior performance of the portfolio on the out-of-sample records. A stability value of 0.04 indicates the absence of a linear fit of the cumulative return with time. The skewness and the kurtosis values exhibit a negatively skewed and platykurtic behavior of the return for the in-sample data. However, the skewness and kurtosis for the out-of-sample records were positive and mesokurtic, respectively. The daily value at risk indicates that with a probability of

0.95, the loss yielded by the portfolio did not exceed 2.35%, 2.38%, and 2.36%, for the in-sample records, out-of-sample records, and all records, respectively, over one day. The alpha values for the portfolio were positive for both in-sample and out-of-sample records indicating that the portfolio consistently outperformed the benchmark NIFTY 50 index. The portfolio yielded an excess return of 0.05% and 0.05% over the in-sample and out-of-sample records, respectively. The beta values were consistently less than 1 indicating a lower volatility of the portfolio in comparison to the benchmark NIFTY 50 index.

**TABLE 2.3.** THE PERFORMANCE OF THE CONSUMER DURABLES SECTOR PORTFOLIO ON THE IN-SAMPLE AND OUT-OF-SAMPLE DATA

| Metric | In-sample data | Out-of-sample data | Overall data |
|---|---|---|---|
| Annual return | 9.51% | 16.92% | 12.79% |
| Cumulative Return | 14.27% | 20.68% | 37.90% |
| Annual volatility | 18.97% | 19.44% | 19.17% |
| Max drawdown | -28.26% | -15.67% | -28.26% |
| Sharpe ratio | 0.57 | 0.90 | 0.72 |
| Calmar ratio | 0.34 | 1.08 | 0.45 |
| Sortino ratio | 0.79 | 1.36 | 1.03 |
| Omega ratio | 1.10 | 1.23 | 1.15 |
| Tail ratio | 0.98 | 1.07 | 0.95 |
| Skewness | -0.53 | 1.26 | 0.31 |
| Kurtosis | 1.84 | 53.78 | 26.51 |
| Stability | 0.31 | 0.00 | 0.04 |
| Daily value at risk | -2.35% | -2.38% | -2.36% |
| Alpha | 0.05 | 0.05 | 0.04 |
| Beta | 0.78 | 0.63 | 0.74 |

*FMCG sector:* Based on the NSE's report published on January 4, 2021, the ten stocks that have the maximum free-float market capitalization in the FMCG sector, and their contributions to the overall index of the sector are as follows: (i) ITC (ITC): 32.42, (ii) Hindustan Unilever (HINDUNILVR): 21.83, (iii) Nestle India (NESTLEIND): 7.96, (iv) Britannia Industries (BRITANNIA): 5.72, (v) Tata Consumer Products (TATACONSUM): 5.66, (vi) Varun

Beverages (VBL): 4.73, (vii) Godrej Consumer Products (GODREJCP): 4.01, (viii) Dabur India (DABUR): 3.45, (ix) United Spirits (MCDOWELL-N): 3.13, and (x) Marico (MARICO): 3.11 (NSE Website). The ticker names of the stocks are mentioned in parentheses. The ticker names serve as the unique identifiers for the stocks on a given stock exchange.

Figure 2.13 shows how the number of shares for the stocks constituting the rebalanced FMCG sector portfolio varied over the entire period (i.e., including both in-sample and out-of-sample records). The initial number of shares for the stocks in the portfolio on January 4, 2021, were as follows: (i) ITC (ITC): 526, (ii) Hindustan Unilever (HINDUNILVR): 43, (iii) Nestle India (NESTLEIND): 5, (iv) Britannia Industries (BRITANNIA): 30, (v) Tata Consumer Products (TATACONSUM): 170, (vi) Varun Beverages (VBL): 503, (vii) Godrej Consumer Products (GODREJCP): 191, (viii) Dabur India (DABUR): 191, (ix) United Spirits (MCDOWELL-N): 171, and (x) Marico (MARICO): 255.

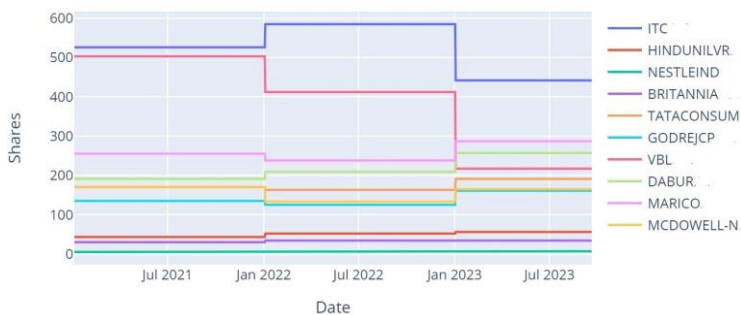

**Figure 2.13.** The daily number of shares of each stock in the FMCG sector portfolio from January 4, 2021, to September 20, 2023.

Figure 2.14 depicts how the weights corresponding to the stocks of the FMCG sector portfolio varied over the entire period. While the daily variation of weights is shown, the rebalancing was done only yearly.

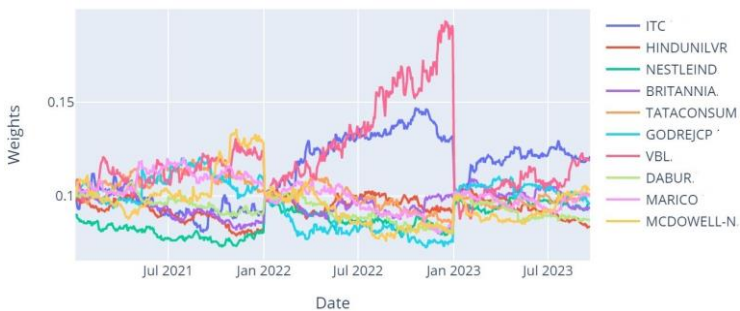

**Figure 2.14.** The daily allocation of weights to each stock of the FMCG sector portfolio from January 4, 2021, to September 20, 2023.

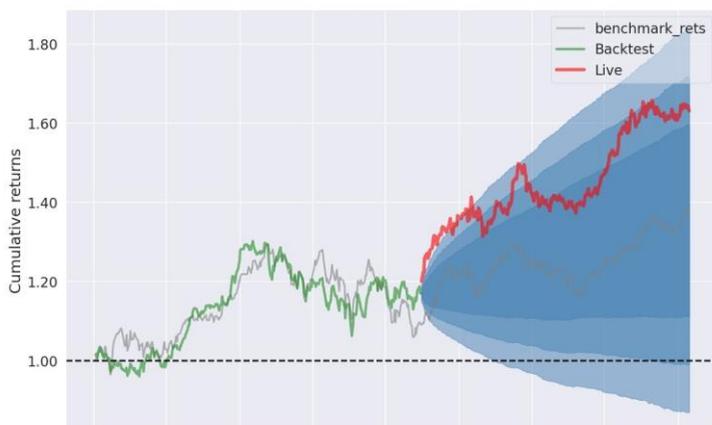

**Figure 2.15.** The cumulative return of the FMCG sector portfolio and the cumulative return of the benchmark index of NIFTY 50 from January 4, 2021, to September 20, 2023. The green, red, and gray lines indicate the cumulative returns for the in-sample records, out-of-sample records of the FMCG sector portfolio, and the benchmark NIFTY 50 index.

Figure 2.15 exhibits the backtesting results of the portfolio performance on its cumulative returns and its comparison with the benchmark cumulative returns of the NIFTY 50 index. The rebalanced

FMCG sector portfolio yielded a higher cumulative return compared to the benchmark NIFTY 50 index most of the time, particularly on the out-of-sample records.

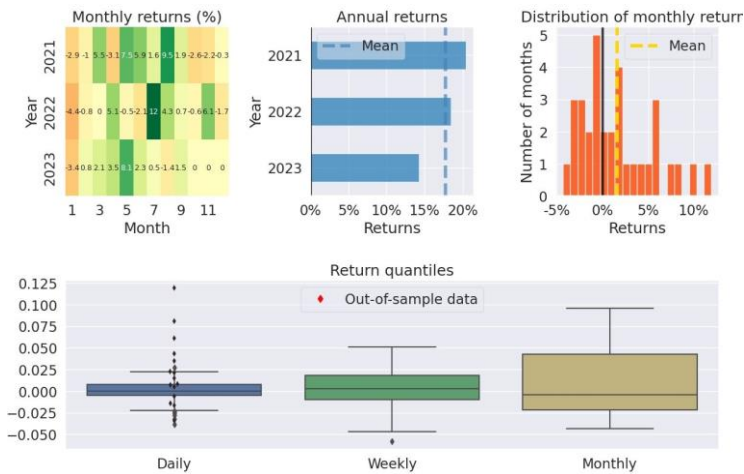

**Figure 2.16.** The statistical distribution and box plots of the daily, weekly, monthly, and annual returns of the rebalanced portfolio of the FMCG sector.

Figure 2.16 depicts several statistical features of the FMCG sector portfolio returns, including the monthly returns of the portfolios over the entire period, the annual returns, the distribution of monthly returns, and the box plots of the daily, monthly, and yearly returns.

The detailed performance results of the rebalanced portfolio of the FMCG sector are presented in Table 2.4. It is observed that the portfolio yielded substantial annual and cumulative returns, while its annual volatility was low. While the values of the Sharpe ratio, Calmar ratio, and Sortino ratio are all less than 1 for the in-sample records, their values are appreciably larger than 1 for the out-of-sample records indicating far superior performance of the portfolio on the out-of-sample records in comparison to the in-sample performance. The tail ratio has been greater than 1 for both in-sample and out-of-sample records indicating that the portfolio yielded positive returns more frequently in comparison to negative returns (i.e., loss). A stability value of 0.87 indicates a good linear fit of the cumulative return with time. The skewness and the kurtosis values exhibit a negatively skewed and platykurtic behavior of the return. The daily value at risk indicates that with a probability of 0.95, the loss yielded by the portfolio did not exceed 1.97%, 1.50%, and 1.77%, for the in-sample records, out-of-sample records, and all records, respectively, over one day. The *alpha* values for the portfolio were positive for both in-sample and out-of-sample records indicating that the portfolio consistently outperformed the benchmark NIFTY 50 index. The *beta* values were less than 1 indicating a lower volatility in comparison to the benchmark NIFTY 50 index.

**TABLE 2.4.** THE PERFORMANCE OF THE FMCG SECTOR PORTFOLIO ON THE IN-SAMPLE AND OUT-OF-SAMPLE DATA

| Metric | In-sample data | Out-of-sample data | Overall data |
|---|---|---|---|
| Annual return | 11.26% | 31.80% | 20.08% |
| Cumulative Return | 16.96% | 39.37% | 63.02% |
| Annual volatility | 16.00% | 12.81% | 14.65% |
| Max drawdown | -18.38% | -8.35% | -18.38% |
| Sharpe ratio | 0.75 | 2.22 | 1.32 |
| Calmar ratio | 0.61 | 3.81 | 1.09 |
| Sortino ratio | 1.07 | 3.56 | 1.96 |
| Omega ratio | 1.13 | 1.44 | 1.24 |
| Tail ratio | 1.06 | 1.33 | 1.08 |
| Skewness | -0.28 | -0.02 | -0.23 |
| Kurtosis | 0.50 | 1.10 | 0.84 |
| Stability | 0.44 | 0.84 | 0.87 |
| Daily value at risk | -1.97% | -1.50% | -1.77% |
| Alpha | 0.07 | 0.18 | 0.12 |
| Beta | 0.60 | 0.61 | 0.60 |

*Information Technology (IT) sector:* As per the report published by the NSE on June 30, 2022, the ten stocks with the largest free-float market capitalization and their respective contributions (in percent) to the overall index of the IT sector are as follows: (i) Infosys (INFY): 27.07, (ii) Tata Consultancy Services (TCS): 25.83, (iii) HCL Technologies (HCLTECH): 9.34, (iv) Tech Mahindra (TECHM): 9.09, (v) Wipro (WIPRO): 8.30, (vi) LTIMindtree (LTIM): 6.93, (vii) Coforge (COFORGE): 4.49, (viii) Persistent Systems (PERSISTENT): 4.27, (ix) MphasiS (MPHASIS): 2.86, and (x) L&T Technology Services (LTTS): 1.82 (NSE Website). The ticker names of the stocks are mentioned in parentheses against their names. The ticker names are the unique identifiers of the stocks in a stock exchange.

Figure 2.17 shows how the number of shares for the stocks constituting the rebalanced *IT* sector portfolio varied over the entire period (i.e., including both in-sample and out-of-sample records). The initial number of shares for the stocks in the portfolio on January 4, 2021, were as follows: (i) Infosys (INFY): 84, (ii) Tata Consultancy Services (TCS): 36, (iii) HCL Technologies (HCLTECH): 117, (iv) Tech Mahindra (TECHM): 115, (v) Wipro (WIPRO): 261, (vi) LTIMindtree (LTIM): 27, (vii) Coforge (COFORGE): 38, (viii) Persistent Systems (PERSISTENT): 68, (ix) MphasiS (MPHASIS): 70, and (x) L&T Technology Services (LTTS): 42.

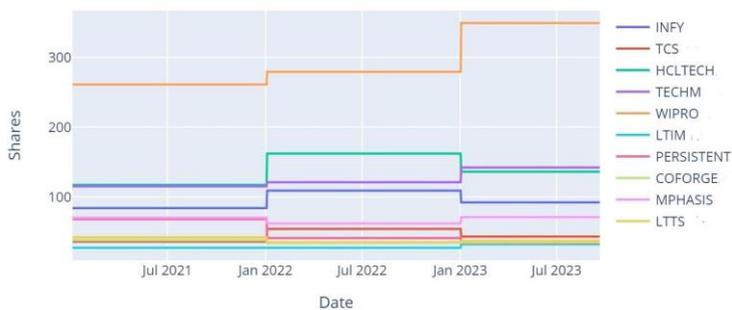

**Figure 2.17.** The daily number of shares of each stock in the IT sector portfolio from January 4, 2021, to September 20, 2023.

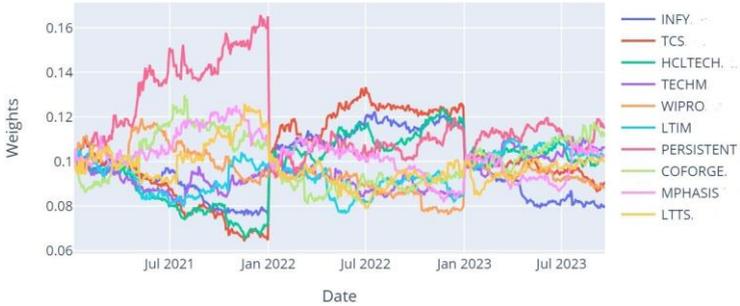

**Figure 2.18.** The daily allocation of weights to each stock of the IT sector portfolio from January 4, 2021, to September 20, 2023.

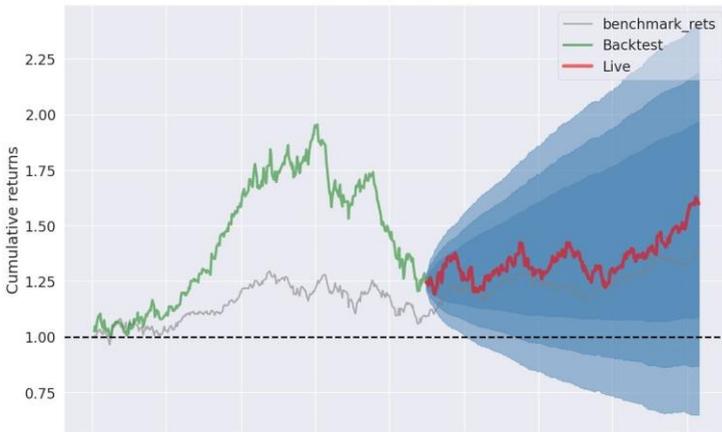

**Figure 2.19.** The cumulative return of the IT sector portfolio and the cumulative return of the benchmark index of NIFTY 50 from January 4, 2021, to September 20, 2023. The green, red, and gray lines indicate the cumulative returns for the in-sample records, out-of-sample records of the IT sector portfolio, and the benchmark NIFTY 50 index.

Figure 2.18 depicts how the weights corresponding to the stocks of the IT sector portfolio varied over the entire period. While the daily variation of weights is shown, the rebalancing was done only yearly.

Figure 2.19 exhibits the backtesting results of the portfolio performance on its cumulative returns and its comparison with the benchmark cumulative returns of the NIFTY 50 index. The rebalanced IT sector portfolio yielded a consistently higher cumulative return compared to the benchmark NIFTY 50 index.

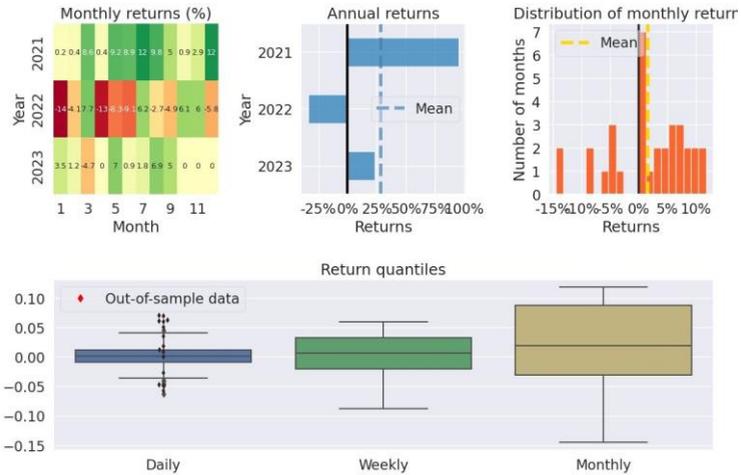

**Figure 2.20.** The statistical distribution and box plots of the daily, weekly, monthly, and annual returns of the rebalanced portfolio of the IT sector.

Figure 2.20 depicts several statistical features of the IT sector portfolio returns, including the monthly returns of the portfolios over the entire period, the annual returns, the distribution of monthly returns, and the box plots of the daily, monthly, and yearly returns.

The detailed performance results of the rebalanced portfolio of the IT sector are presented in Table 2.5. It is observed that the portfolio yielded substantial annual and cumulative returns, while its annual volatility was moderate. While the values of the Sharpe ratio, Calmar ratio, and Sortino ratio are all less than 1 for the in-sample records, their values are larger than 1 for the out-of-sample records indicating a superior performance of the portfolio on the out-of-sample records in comparison to the in-sample performance. A stability value of 0.03 indicates the absence of a linear fit of the cumulative return with time. The skewness and the kurtosis values exhibit a negatively skewed and platykurtic behavior of the return. The daily value at risk indicates that with a probability of 0.95, the loss yielded by the portfolio did not exceed 3.23%, 2.48%, and 2.91%, for the in-sample records, out-of-sample records, and all records, respectively, over one day. The alpha values for the portfolio were positive for both in-sample and out-of-sample records indicating that the portfolio consistently outperformed the benchmark NIFTY 50 index. The *beta* values of 0.95 for the in-sample records indicate the portfolio exhibited marginally less volatility in comparison to the benchmark NIFTY 50 index. However, for the out-of-sample records, the volatility was higher in comparison to the benchmark NIFTY 50 as indicated by the *beta* value of 1.15.

**TABLE 2.5.** THE PERFORMANCE OF THE IT SECTOR PORTFOLIO ON THE IN-SAMPLE AND OUT-OF-SAMPLE DATA

| Metric | In-sample data | Out-of-sample data | Overall data |
|---|---|---|---|
| Annual return | 16.12% | 23.03% | 19.18% |
| Cumulative Return | 24.54% | 28.30% | 59.79% |
| Annual volatility | 26.24% | 20.37% | 23.76% |
| Max drawdown | -38.32% | -14.14% | -39.09% |
| Sharpe ratio | 0.70 | 1.12 | 0.86 |
| Calmar ratio | 0.42 | 1.63 | 0.49 |
| Sortino ratio | 0.98 | 1.67 | 1.23 |
| Omega ratio | 1.12 | 1.21 | 1.15 |
| Tail ratio | 1.03 | 1.15 | 1.04 |
| Skewness | -0.34 | -0.08 | -0.28 |
| Kurtosis | 0.60 | 0.95 | 0.90 |
| Stability | 0.43 | 0.50 | 0.03 |
| Daily value at risk | -3.23% | -2.48% | -2.91% |
| Alpha | 0.10 | 0.01 | 0.07 |
| Beta | 0.95 | 1.15 | 1.00 |

*Metal sector:* As per the report published by the NSE on January 4, 2021, the ten stocks of the *metal* sector that have the largest free-float market capitalization and their respective contributions (in

percent) to the overall index of the *metal* sector are as follows: (i) Tata
Steel (TATASTEEL): 21.25, (ii) Adani Enterprises (ADANIENT):
16.31, (iii) JSW Steel (JSWSTEEL): 14.80, (iv) Hindalco Industries
(HINDALCO): 14.71, (v) APL Apollo Tubes (APLAPOLLO): 5.80,
(vi) Vedanta (VEDL): 5.41, (vii) Jindal Steel & Power
(JINDALSTEL): 5.26, (viii) NMDC (NMDC): 3.46, (ix) Jindal
Stainless (JSL): 3.36, and (x) Steel Authority of India (SAIL): 2.77
(NSE Website).

Figure 2.21 shows how the number of shares for the stocks
constituting the rebalanced *metal* sector portfolio varied over the entire
period (i.e., including both in-sample and out-of-sample records). The
initial number of shares for the stocks in the portfolio on January 4,
2021, were as follows: (i) Tata Steel (TATASTEEL): 3693, (ii) Adani
Enterprises (ADANIENT): 204, (iii) JSW Steel (JSWSTEEL): 279,
(iv) Hindalco Industries (HINDALCO): 429, (v) APL Apollo Tubes
(APLAPOLLO): 234, (vi) Vedanta (VEDL): 1099, (vii) Jindal Steel
& Power (JINDALSTEL): 373, (viii) NMDC (NMDC): 1067, (ix)
Jindal Stainless (JSL): 1324, and (x) Steel Authority of India (SAIL):
1537.

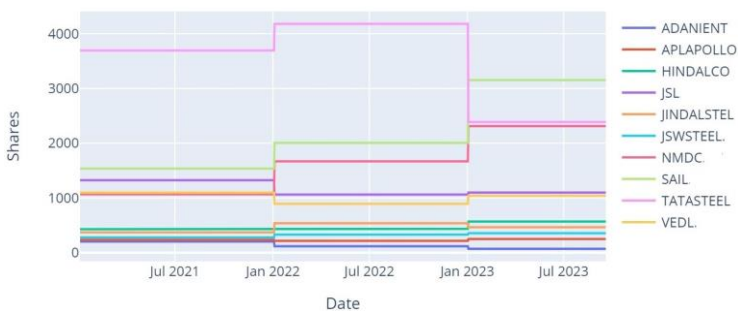

**Figure 2.21.** The daily number of shares of each stock in the *metal*
sector portfolio from January 4, 2021, to September 20, 2023.

Figure 2.22 depicts how the weights corresponding to the stocks of
the *metal* sector portfolio varied over the entire period. While the daily
variation of weights is shown, the rebalancing was done only yearly.

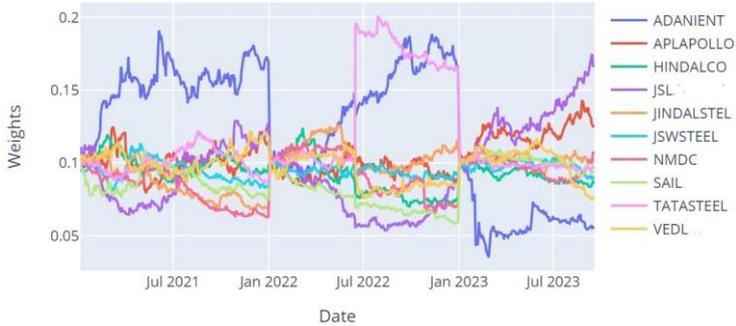

**Figure 2.22.** The daily allocation of weights to each stock of the *metal* sector portfolio from January 4, 2021, to September 20, 2023.

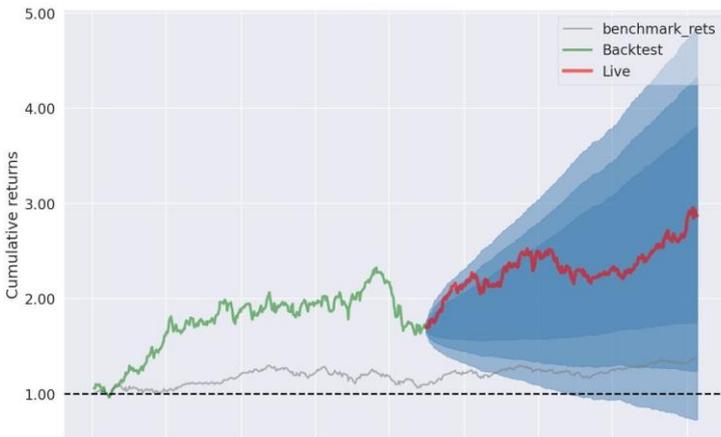

**Figure 2.23.** The cumulative return of the *metal* sector portfolio and the cumulative return of the benchmark index of NIFTY 50 from January 4, 2021, to September 20, 2023. The green, red, and gray lines indicate the cumulative returns for the in-sample records, out-of-sample records of the *metal* sector portfolio, and the benchmark NIFTY 50 index.

Figure 2.23 exhibits the backtesting results of the portfolio performance on its cumulative returns and its comparison with the benchmark cumulative returns of the NIFTY 50 index. The rebalanced *metal* sector portfolio yielded a consistently higher cumulative return compared to the benchmark NIFTY 50 index.

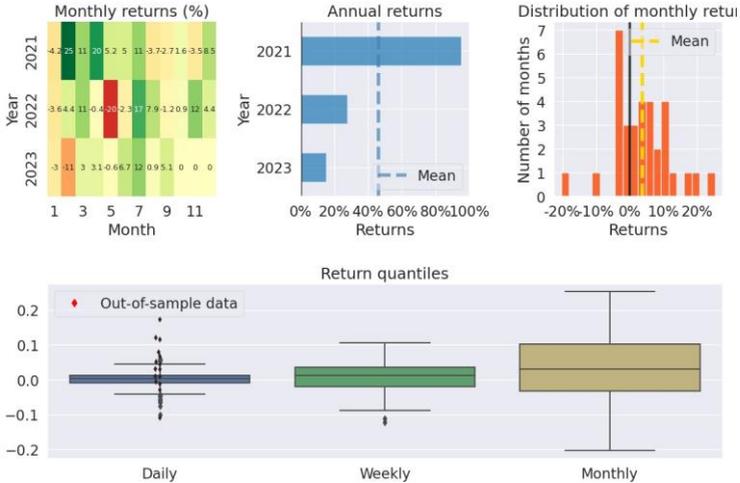

**Figure 2.24.** The statistical distribution and box plots of the daily, weekly, monthly, and annual returns of the rebalanced portfolio of the *metal* sector.

Figure 2.24 depicts several statistical features of the *metal* sector portfolio returns, including the monthly returns of the portfolios over the entire period, the annual returns, the distribution of monthly returns, and the box plots of the daily, monthly, and yearly returns.

The detailed performance results of the rebalanced portfolio of the *metal* sector are presented in Table 2.6. It is observed that the portfolio yielded substantial annual and cumulative returns, while its annual volatility was a little high, particularly on the in-sample records. The values of the Sharpe ratio, Sortino ratio, Calmar ratio, and Omega ratio for both in-sample and out-of-sample are all greater than 1 indicating a good performance, particularly over the out-of-sample records (i.e., during the portfolio test period). The tail ratio was also greater than 1 except for the in-sample records implying that the portfolio yielded positive returns more frequently than negative returns (i.e., loss). A stability value of 0.73 indicates a reasonably good linear fit of the cumulative return with time. The skewness and the kurtosis values exhibit a negatively skewed and platykurtic behavior of the return. The daily value at risk indicates that with a probability of 0.95, the loss yielded by the portfolio did not exceed 4.15%, 2.48%, and 3.49%, for the in-sample records, out-of-sample records, and all records, respectively, over one day. The *alpha* values for the portfolio were positive for both in-sample and out-of-sample records indicating that the portfolio consistently outperformed the benchmark NIFTY 50 index. The portfolio yielded an excess return of 0.34% and 0.27% over the in-sample and out-of-sample records, respectively. The *beta* values were greater than 1 for both in-sample and out-of-sample records indicating a higher volatility of the portfolio in comparison to volatility of the benchmark NIFTY 50 index.

**TABLE 2.6.** THE PERFORMANCE OF THE METAL SECTOR PORTFOLIO ON THE IN-SAMPLE AND OUT-OF-SAMPLE DATA

| **Metric** | **In-sample data** | **Out-of-sample data** | **Overall data** |
|---|---|---|---|
| Annual return | 42.91% | 53.18% | 47.44% |
| Cumulative Return | 68.92% | 66.71% | 181.60% |
| Annual volatility | 34.28% | 21.10% | 29.09% |
| Max drawdown | -30.26% | -14.44% | -30.26% |
| Sharpe ratio | 1.21 | 2.13 | 1.48 |
| Calmar ratio | 1.42 | 3.68 | 1.57 |
| Sortino ratio | 1.69 | 3.29 | 2.11 |
| Omega ratio | 1.23 | 1.42 | 1.29 |
| Tail ratio | 0.98 | 1.03 | 1.18 |
| Skewness | -0.63 | -0.13 | -0.59 |
| Kurtosis | 2.02 | 0.81 | 2.80 |
| Stability | 0.56 | 0.62 | 0.73 |
| Daily value at risk | -4.15% | -2.48% | -3.49% |
| Alpha | 0.34 | 0.27 | 0.30 |
| Beta | 1.18 | 1.14 | 1.17 |

*Pharma sector:* As per the report published by the NSE on January 4, 2021, the ten stocks with the largest free-float market capitalization and their respective contributions to the overall index of the *pharma* sector are as follows: (i) Sun Pharmaceuticals Industries (SUNPHARMA): 24.22, (ii) Dr. Reddy's Labs (DRREDDY): 13.17, (iii) Cipla (CIPLA): 12.05, (iv) Divi's Laboratories (DIVISLAB): 9.29, (v) Lupin (LUPIN): 5.47, (vi) Aurobindo Pharma (AUROPHARMA): 4.98, (vii) Alkem Laboratories (ALKEM): 3.42, (viii) Torrent Pharmaceuticals (TORNTPHARM): 3.41, (ix) Zydus Lifesciences (ZYDUSLIFE): 3.01, and (x) Laurus Labs (LAURUSLABS): 3.01 (NSE Website). The ticker names of the stocks are mentioned in parentheses. The ticker names are the unique identifiers of the stocks listed on a stock exchange.

Figure 2.25 shows how the number of shares for the stocks of the rebalanced *pharma* sector portfolio varied over the entire period (i.e., including both in-sample and out-of-sample records). The initial number of shares for the stocks in the portfolio on January 4, 2021, were as follows: (i) Sun Pharmaceuticals Industries (SUNPHARMA): 173, (ii) Dr. Reddy's Labs (DRREDDY): 19, (iii) Cipla (CIPLA): 124, (iv) Divi's Laboratories (DIVISLAB): 26, (v) Lupin (LUPIN): 101, (vi) Aurobindo Pharma (AUROPHARMA): 110, (vii) Alkem Laboratories (ALKEM): 35, (viii) Torrent Pharmaceuticals (TORNTPHARM), (ix) Zydus Lifesciences (ZYDUSLIFE): 217, and (x) Laurus Labs (LAURUSLABS): 286.

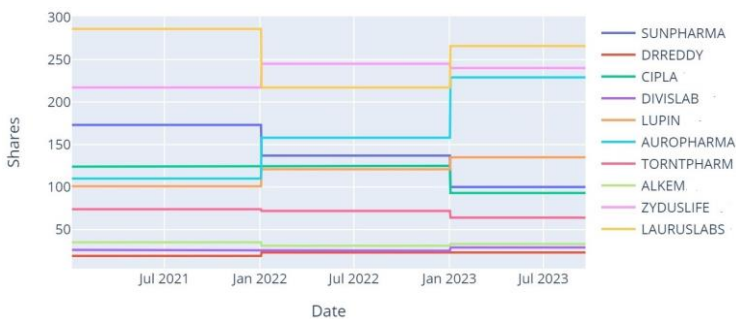

**Figure 2.25.** The daily number of shares of each stock in the *pharma* sector portfolio from January 4, 2021, to September 20, 2023.

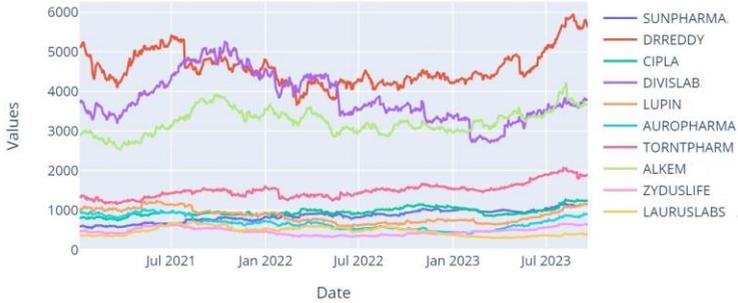

**Figure 2.26.** The daily allocation of weights to each stock of the *pharma* sector portfolio from January 4, 2021, to September 20, 2023.

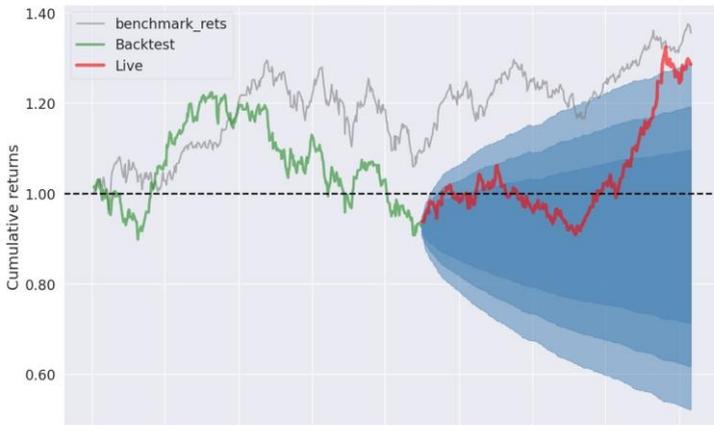

**Figure 2.27.** The cumulative return of the *pharma* sector portfolio and the cumulative return of the benchmark index of NIFTY 50 from January 4, 2021, to September 20, 2023. The green, red, and gray lines indicate the cumulative returns for the in-sample records, out-of-sample records of the *pharma* sector portfolio, and the benchmark NIFTY 50 index.

Figure 2.26 depicts how the weights corresponding to the stocks of the *pharma* sector portfolio varied over the entire period. While the daily variation of weights is shown, the rebalancing was done only yearly.

Figure 2.27 exhibits the backtesting results of the portfolio performance on its cumulative returns and its comparison with the benchmark cumulative returns of the NIFTY 50 index. The rebalanced *pharma* sector portfolio yielded a consistently lower cumulative return compared to the benchmark NIFTY 50 index except for a brief period on the in-sample records.

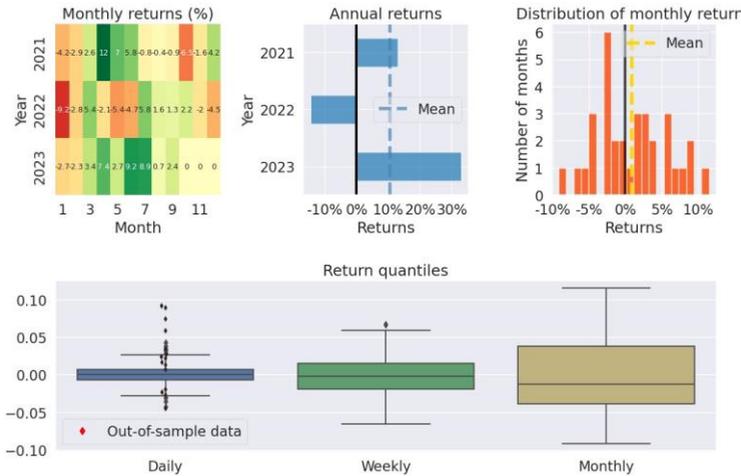

**Figure 2.28.** The statistical distribution and box plots of the daily, weekly, monthly, and annual returns of the rebalanced portfolio of the *pharma* sector.

Figure 2.28 depicts several statistical features of the *pharma* sector portfolio returns, including the monthly returns of the portfolios over the entire period, the annual returns, the distribution of monthly returns, and the box plots of the daily, monthly, and yearly returns.

The detailed performance results of the rebalanced portfolio of the *pharma* sector are presented in Table 2.7. It is observed that while the portfolio yielded negative annual and cumulative returns for the in-sample records, it produced high return and cumulative returns over the out-of-sample records. The annual volatility was low for both in-sample and out-of-sample records. The values of the Sharpe ratio, Sortino ratio, Calmar ratio, Omega ratio, and Tail ratio for the in-sample records were very poor. However, all these ratios exhibited far improved values on the out-of-sample records indicating higher risk-adjusted returns from the portfolio. A stability value of 0.00 indicates that there is no linear fit of the cumulative return with time. The skewness and the kurtosis values exhibit a slightly positively skewed and platykurtic behavior of the return. The daily value at risk indicates that with a probability of 0.95, the loss yielded by the portfolio did not exceed 2.41%, 1.67%, and 2.10%, for the in-sample records, out-of-sample records, and all records, respectively, over one day.

**TABLE 2.7.** THE PERFORMANCE OF THE PHARMA SECTOR PORTFOLIO ON THE IN-SAMPLE AND OUT-OF-SAMPLE DATA

| **Metric** | **In-sample data** | **Out-of-sample data** | **Overall data** |
|---|---|---|---|
| Annual return | -5.00 | 31.35% | 9.89% |
| Cumulative Return | -7.25% | 38.66% | 28.60% |
| Annual volatility | 19.01% | 14.18% | 17.03% |
| Max drawdown | -25.85% | -14.38% | -25.85% |
| Sharpe ratio | -0.17 | 2.00 | 0.64 |
| Calmar ratio | -0.19 | 2.18 | 0.38 |
| Sortino ratio | -0.24 | 3.40 | 0.94 |
| Omega ratio | 0.97 | 1.39 | 1.11 |
| Tail ratio | 0.93 | 1.29 | 1.10 |
| Skewness | 0.02 | 0.55 | 0.11 |
| Kurtosis | 0.85 | 0.77 | 1.15 |
| Stability | 0.02 | 0.46 | 0.00 |
| Daily value at risk | -2.41% | -1.67% | -2.10% |
| Alpha | -0.08 | 0.22 | 0.03 |
| Beta | 0.61 | 0.44 | 0.57 |

The *alpha* values for the portfolio were negative for the in-sample, and positive for out-of-sample records indicating that while the portfolio yielded a lower return in comparison to the return of the benchmark NIFTY 50 over the in-sample records, it yielded an excess return over the benchmark for the out-of-sample records. The *beta* values have been consistently less than 1, which indicates a lower volatility of the portfolio in comparison to the benchmark.

*Private Banks sector:* The report published by the NSE on January 4, 2021, identified the top ten stocks in the *private banks* sector with the largest free-float market capitalization. These stocks and their respective contributions (in percent) to the overall index of the sector are as follows: (i) HDFC Bank (HDFCBANK): 25.99, (ii) ICICI Bank (ICICIBANK): 25.22, (iii) Axis Bank (AXISBANK): 10.88, (iv) IndusInd Bank (INDUSINDBK): 10.68, (v) Kotak Mahindra Bank (KOTAKBANK): 10.47, (vi) Federal Bank (FEDERALBNK): 5.40, (vii) IDFC First Bank (IDFCFIRSTB): 4.70, (viii) Bandhan Bank (BANDHANBNK): 2.91, (ix) RBL Bank (RBLBANK): 2.34, and (x) City Union Bank (CUB): 1.42 (NSE Website). The ticker names of the stocks are mentioned in parentheses. The ticker names are the unique identifiers for the stocks listed on a stock exchange.

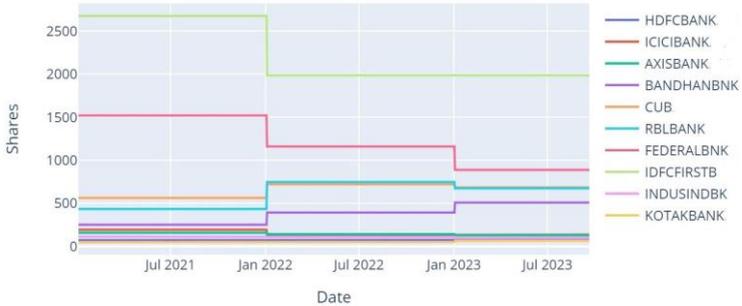

**Figure 2.29.** The daily number of shares of each stock in the *private banks* sector portfolio from January 4, 2021, to September 20, 2023.

Figure 2.29 shows how the number of shares for the stocks constituting the rebalanced *private banks* sector portfolio varied over the entire period (i.e., including both in-sample and out-of-sample records). The initial number of shares for the stocks in the portfolio on January 4, 2021, were as follows: (i) HDFC Bank (HDFCBANK): 72, (ii) ICICI Bank (ICICIBANK): 194, (iii) Axis Bank (AXISBANK): 160, (iv) IndusInd Bank (INDUSINDBK): 113, (v) Kotak Mahindra Bank (KOTAKBANK): 50, (vi) Federal Bank (FEDERALBNK):

1519, (vii) IDFC First Bank (IDFCFIRSTB): 2673, (viii) Bandhan Bank (BANDHANBNK): 252, (ix) RBL Bank (RBLBANK): 435, and (x) City Union Bank (CUB): 563.

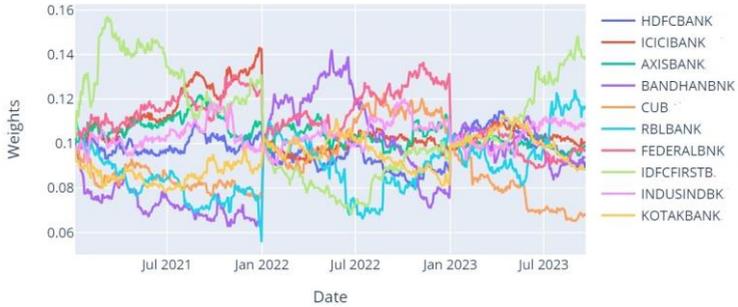

**Figure 2.30.** The daily allocation of weights to each stock of the *private banks* sector portfolio from January 4, 2021, to September 20, 2023.

Figure 2.30 depicts how the weights corresponding to the stocks of the *private banks* sector portfolio varied over the entire period. While the daily variation of weights is shown, the rebalancing was done only yearly.

Figure 2.31 exhibits the backtesting results of the portfolio performance on its cumulative returns and its comparison with the benchmark cumulative returns of the NIFTY 50 index. The rebalanced *private banks* sector portfolio yielded a consistently lower cumulative return compared to the benchmark NIFTY 50 index except for a very brief period in the in-sample records.

Figure 2.32 depicts several statistical features of the *private banks* sector portfolio returns, including the monthly returns of the portfolios over the entire period, the annual returns, the distribution of monthly returns, and the box plots of the daily, monthly, and yearly returns.

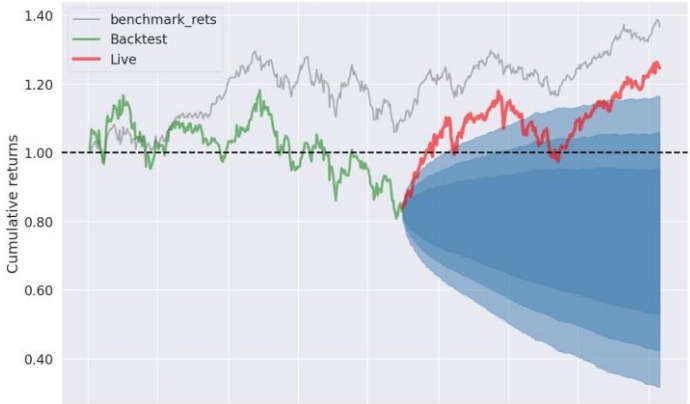

**Figure 2.31.** The cumulative return of the *private banks* sector portfolio and the cumulative return of the benchmark index of NIFTY 50 from January 4, 2021, to September 20, 2023. The green, red, and gray lines indicate the cumulative returns for the in-sample records, out-of-sample records of the *private banks* sector portfolio, and the benchmark NIFTY 50 index.

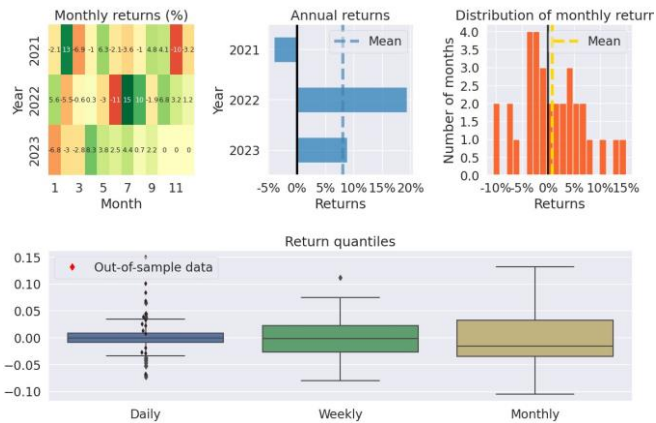

**Figure 2.32.** The statistical distribution and box plots of the daily, weekly, monthly, and annual returns of the rebalanced portfolio of the *private banks* sector.

| Metric | In-sample data | Out-of-sample data | Overall data |
|---|---|---|---|
| Annual return | -12.11% | 40.56% | 8.58% |
| Cumulative return | -17.27% | 50.59% | 24.59% |
| Annual volatility | 26.12% | 17.97% | 22.84% |
| Max drawdown | -31.67% | -17.40% | -31.67% |
| Sharpe ratio | -0.36 | 1.99 | 0.48 |
| Calmar ratio | -0.38 | 2.33 | 0.27 |
| Sortino ratio | -0.49 | 3.01 | 0.66 |
| Omega ratio | 0.94 | 1.39 | 1.09 |
| Tail ratio | 1.01 | 1.06 | 0.99 |
| Skewness | -0.41 | -0.17 | -0.44 |
| Kurtosis | 2.44 | 1.07 | 2.89 |
| Stability | 0.48 | 0.52 | 0.09 |
| Daily value at risk | -3.33% | -2.12% | -2.84% |
| Alpha | -0.18 | 0.14 | -0.05 |
| Beta | 1.22 | 1.16 | 1.21 |

The detailed performance results of the rebalanced portfolio of the *private banks* sector are presented in Table 2.8. It is observed that while the portfolio yielded negative annual and cumulative returns for the in-sample records, it produced high return and cumulative returns over the out-of-sample records. The annual volatility was moderate for both in-sample and out-of-sample records. The values of the Sharpe ratio, Sortino ratio, Calmar ratio, and Omega ratio for the in-sample records were very poor. However, all these ratios exhibited improved values on the out-of-sample records indicating higher risk-adjusted returns from the portfolio. The Tail ratio, however, was higher than 1 for both in-sample and out-of-sample records indicating that the portfolio yielded positive returns more frequently than negative returns. A very low value of 0.09 for stability indicates that there is no linear fit of the cumulative return with time. The skewness and the kurtosis values exhibit a slightly negatively skewed and platykurtic behavior of the return. The daily value at risk indicates that with a probability of 0.95, the loss yielded by the portfolio did not exceed

3.33%, 2.12%, and 2.84%, for the in-sample records, out-of-sample records, and all records, respectively, over one day. The *alpha* values for the portfolio were negative for the in-sample, and positive for out-of-sample records indicating that while the portfolio yielded a lower return in comparison to the return of the benchmark NIFTY 50 over the in-sample records, it yielded an excess return over the benchmark for the out-of-sample records. The *beta* values have been consistently higher than 1, which indicates a higher volatility of the portfolio in comparison to the benchmark.

*PSU Banks sector:* As per the report published by the NSE on January 4, 2021, the ten stocks that have the largest free-float market capitalization in this sector, and their respective contributions (in percent) to the overall index of the sector are as follows: (i) State Bank of India (SBIN): 31.64, (ii) Bank of Baroda (BANKBARODA): 18.34, (iii) Canara Bank (CANBK): 11.63, (iv) Punjab National Bank (PNB): 11.56, (v) Union Bank of India (UNIONBANK): 9.38, (vi) Indian Bank (INDIANB): 5.46, (vii) Bank of India (BANKINDIA): 4.42, (viii) Bank of Maharashtra (MAHABANK): 2.45, (ix) Indian Overseas Bank (IOB): 1.83, and (x) Central Bank of India (CENTRALBK): 1.61 (NSE Website). The ticker names of the stocks, which are their unique identifiers, are mentioned in parentheses.

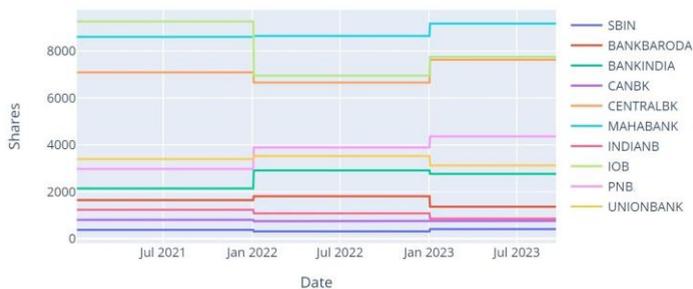

**Figure 2.33.** The daily number of shares of each stock in the *PSU banks* sector portfolio from January 4, 2021, to September 20, 2023.

Figure 2.33 shows how the number of shares for the stocks constituting the rebalanced *PSU banks* sector portfolio varied over the entire period (i.e., including both in-sample and out-of-sample records). The initial number of shares for the stocks in the portfolio on

January 4, 2021, were as follows: (i) State Bank of India (SBIN): 373, (ii) Bank of Baroda (BANKBARODA): 1644, (iii) Canara Bank (CANBK): 806, (iv) Punjab National Bank (PNB): 2974, (v) Union Bank of India (UNIONBANK): 3399, (vi) Indian Bank (INDIANB): 1234, (vii) Bank of India (BANKINDIA): 2140, (viii) Bank of Maharashtra (MAHABANK): 8605, (ix) Indian Overseas Bank (IOB): 9259, and (x) Central Bank of India (CENTRALBK): 7092.

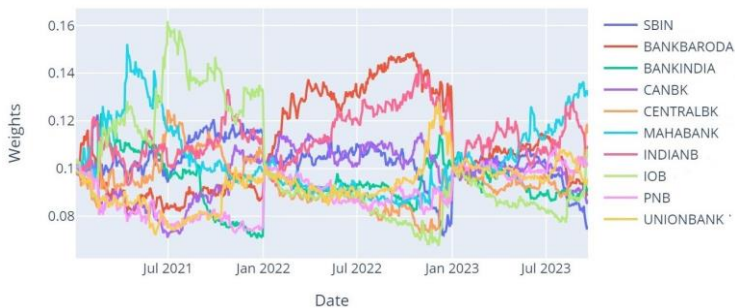

**Figure 2.34.** The daily allocation of weights to each stock of the *PSU banks* sector portfolio from January 4, 2021, to September 20, 2023.

Figure 2.34 depicts how the weights corresponding to the stocks of the *PSU banks* sector portfolio varied over the entire period. While the daily variation of weights is shown, the rebalancing was done only yearly.

Figure 2.35 exhibits the backtesting results of the portfolio performance on its cumulative returns and its comparison with the benchmark cumulative returns of the NIFTY 50 index. The rebalanced *PSU banks* sector portfolio yielded a consistently higher cumulative return compared to the benchmark NIFTY 50 index.

Figure 2.36 depicts several statistical features of the *PSU banks* sector portfolio returns, including the monthly returns of the portfolios over the entire period, the annual returns, the distribution of monthly returns, and the box plots of the daily, monthly, and yearly returns.

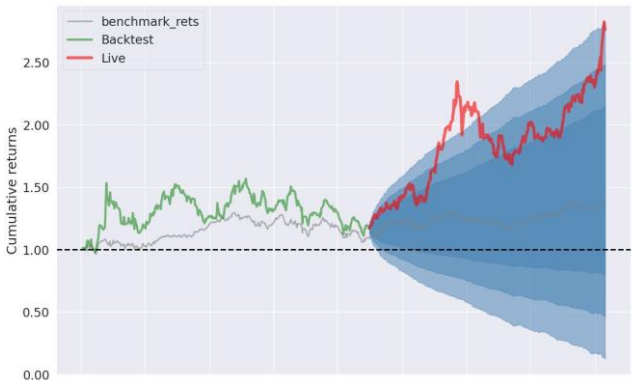

**Figure 2.35.** The cumulative return of the *PSU banks* sector portfolio and the cumulative return of the benchmark index of Nifty 50 from January 4, 2021, to September 20, 2023. The green, red, and gray lines indicate the cumulative returns for the in-sample records, out-of-sample records of the *PSU banks* sector portfolio, and the benchmark NIFTY 50 index.

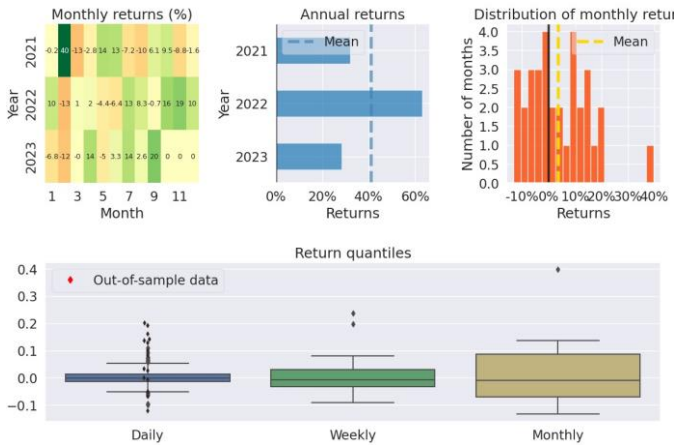

**Figure 2.36.** The statistical distribution and box plots of the daily, weekly, monthly, and annual returns of the rebalanced portfolio of the *PSU banks* sector.

 THE PERFORMANCE OF THE PSU BANKS SECTOR
PORTFOLIO ON THE IN-SAMPLE AND OUT-OF-SAMPLE DATA

| Metric | In-sample data | Out-of-sample data | Overall data |
|---|---|---|---|
| Annual return | 11.02% | 104.95% | 46.31% |
| Cumulative return | 16.58% | 136.99% | 176.28% |
| Annual volatility | 37.97% | 30.90% | 34.98% |
| Max drawdown | -29.25% | -28.30% | -29.25% |
| Sharpe ratio | 0.46 | 2.48 | 1.26 |
| Calmar ratio | 0.38 | 3.71 | 1.58 |
| Sortino ratio | 0.71 | 4.09 | 1.98 |
| Omega ratio | 1.09 | 1.54 | 1.25 |
| Tail ratio | 1.08 | 1.34 | 1.19 |
| Skewness | 0.33 | 0.21 | 0.27 |
| Kurtosis | 2.95 | 2.28 | 2.96 |
| Stability | 0.03 | 0.70 | 0.66 |
| Daily value at risk | -4.71% | -3.59% | -4.23% |
| Alpha | 0.08 | 0.66 | 0.32 |
| Beta | 1.18 | 1.35 | 1.22 |

The detailed performance results of the rebalanced portfolio of the *PSU banks* sector are presented in Table 2.9. It is observed that the portfolio yielded very high annual and cumulative returns, particularly on the out-of-sample records. The annual volatility of the portfolio was also high for both in-sample and out-of-sample records. While the values of the Sharpe ratio, Sortino ratio, and Calmar ratio are poor for the in-sample records, the ratios are quite impressive for the out-of-sample records. Hence, the portfolio yielded high risk-adjusted returns on the out-of-sample records. The Omega ratio and Tail ratio for both in-sample and out-of-sample are all greater than 1 indicating that the portfolio had its risk effectively covered by its return and the number of cases it produced positive return was higher than the number of cases its return was negative. A stability value of 0.66 indicates a modest linear fit of the cumulative return with time. The skewness and the kurtosis values exhibit a positively skewed and nearly mesokurtic behavior of the return. The daily value at risk indicates that with a probability of 0.95, the loss yielded by the portfolio did not exceed

4.71%, 3.59%, and 4.23%, for the in-sample records, out-of-sample records, and all records, respectively, over one day. The *alpha* values for the portfolio were positive for both in-sample and out-of-sample records indicating that the portfolio consistently outperformed the benchmark NIFTY 50 index. The portfolio yielded an excess return of 0.08% and 0.66% over the in-sample and out-of-sample records, respectively. The *beta* values were consistently higher than 1, indicating a higher volatility of the portfolio in comparison to the benchmark.

***Realty sector:*** As per the report published by the NSE on January 4, 2021, the ten stocks that have the largest free-float market capitalization in the *realty* sector and their contributions (in percent) to the overall index of the sector are as follows: (i) DLF (DLF): 26.42, (ii) Macrotech Developers (LODHA): 15.51. (iii) Godrej Properties (GODREJPROP): 14.25, (iv) Phoenix Mills (PHOENIXLTD): 13.55, (v) Oberoi Realty (OBEROIRLTY): 10.80, (vi) Prestige Estate Projects (PRESTIGE): 6.80, (vii) Brigade Enterprises (BRIGADE): 5.69, (viii) Mahindra Lifespace Developers (MAHLIFE): 3.29, (ix) Sobha (SOBHA): 2.05, and (x) Indiabulls Real Estate (IBREALEST): 1.65 (NSE Website). The ticker names of the stocks, which are their unique identifiers, are mentioned in parentheses.

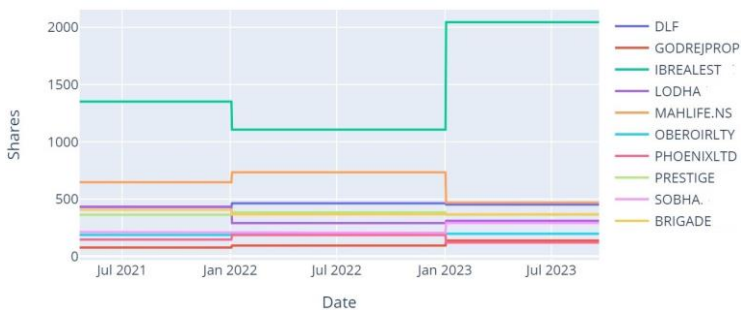

**Figure 2.37.** The daily no of shares of each stock in the realty sector portfolio from January 4, 2021, to September 20, 2023.

Figure 2.37 shows how the number of shares for the stocks constituting the rebalanced *realty* sector portfolio varied over the entire period (i.e., including both in-sample and out-of-sample records). The initial number of shares for the stocks in the portfolio on January 4, 2021, were as follows: (i) DLF (DLF): 431, (ii) Macrotech Developers (LODHA): 431, (iii) Godrej Properties (GODREJPROP): 76, (iv) Phoenix Mills (PHOENIXLTD): 146, (v) Oberoi Realty (OBEROIRLTY): 187, (vi) Prestige Estate Projects (PRESTIGE): 362, (vii) Brigade Enterprises (BRIGADE): 407, (viii) Mahindra Lifespace Developers (MAHLIFE): 646, (ix) Sobha (SOBHA): 211, and (x) Indiabulls Real Estate (IBREALEST): 1352.

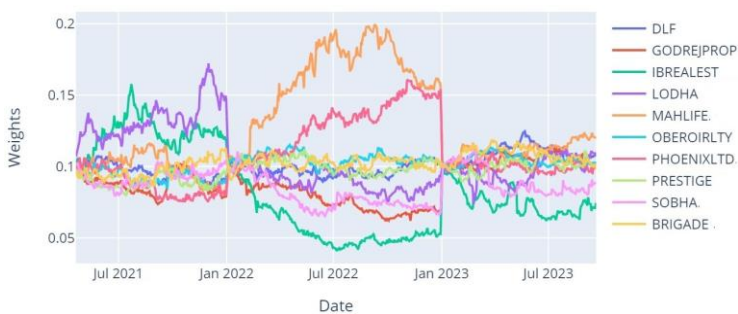

**Figure 2.38.** The daily allocation of weights to each stock of the realty sector portfolio from January 4, 2021, to September 20, 2023.

Figure 2.38 depicts how the weights corresponding to the stocks of the auto sector portfolio varied over the entire period. While the daily variation of weights is shown, the rebalancing was done only yearly.

Figure 2.39 exhibits the backtesting results of the portfolio performance on its cumulative returns and its comparison with the benchmark cumulative returns of the NIFTY 50 index. The rebalanced *realty* sector portfolio yielded a consistently higher cumulative return compared to the benchmark NIFTY 50 index.

Figure 2.40 depicts several statistical features of the *realty* sector portfolio returns, including the monthly returns of the portfolios over the entire period, the annual returns, the distribution of monthly returns, and the box plots of the daily, monthly, and yearly returns.

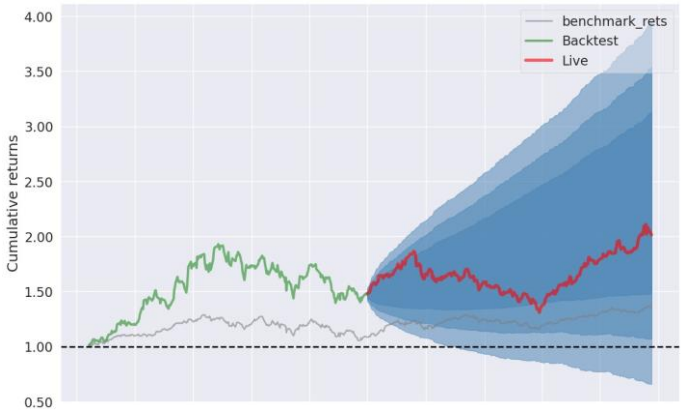

**Figure 2.39.** The cumulative return of the *realty* sector portfolio and the cumulative return of the benchmark index of NIFTY 50 from January 4, 2021, to September 20, 2023. The green, red, and gray lines indicate the cumulative returns for the in-sample records, out-of-sample records of the *realty* sector portfolio, and the benchmark NIFTY 50 index.

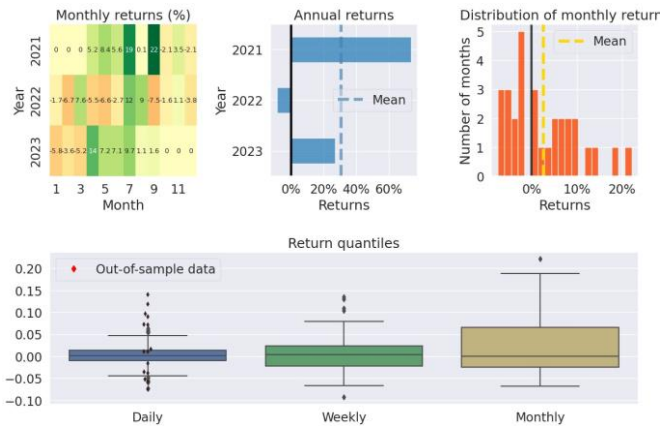

**Figure 2.40.** The statistical distribution and box plots of the daily, weekly, monthly, and annual returns of the rebalanced portfolio of the *realty* sector.

**TABLE 2.10.** THE PERFORMANCE OF THE REALTY SECTOR
PORTFOLIO ON THE IN-SAMPLE AND OUT-OF-SAMPLE DATA

| Metric | In-sample data | Out-of-sample data | Overall data |
|---|---|---|---|
| Annual return | 38.40% | 30.08% | 34.16% |
| Cumulative return | 47.05% | 37.05% | 101.53% |
| Annual volatility | 32.65% | 19.06% | 26.68% |
| Max drawdown | -27.67% | -29.74% | -32.12% |
| Sharpe ratio | 1.16 | 1.48 | 1.24 |
| Calmar ratio | 1.39 | 1.01 | 1.06 |
| Sortino ratio | 1.74 | 2.15 | 1.84 |
| Omega ratio | 1.21 | 1.27 | 1.24 |
| Tail ratio | 1.22 | 1.06 | 1.09 |
| Skewness | -0.04 | -0.32 | -0.07 |
| Kurtosis | 1.03 | 1.08 | 2.01 |
| Stability | 0.32 | 0.12 | 0.24 |
| Daily value at risk | -3.96% | -2.29% | -3.23% |
| Alpha | 0.30 | 0.11 | 0.18 |
| Beta | 1.27 | 0.96 | 1.17 |

The detailed performance results of the rebalanced portfolio of the *realty* sector are presented in Table 2.10. It is observed that the portfolio yielded high annual and cumulative returns, while its annual volatility for the in-sample records was slightly high. The values of the Sharpe ratio, Sortino ratio, Calmar ratio, Omega ratio, and Tail ratio for both in-sample and out-of-sample records are all greater than 1 indicating a good performance, particularly over the out-of-sample records (i.e., during the portfolio test period). A low stability value of 0.24 indicates the absence of a linear fit of the cumulative return with time. The skewness and the kurtosis values exhibit a slightly negatively skewed and platykurtic behavior of the return. The daily value at risk indicates that with a probability of 0.95, the loss yielded by the portfolio did not exceed 3.96%, 2.29%, and 3.23%, for the in-sample records, out-of-sample records, and all records, respectively, over one day. The alpha values for the portfolio were positive for both in-sample and out-of-sample records indicating that the portfolio consistently yielded a higher return compared to the return produced by the benchmark NIFTY 50 index. The portfolio yielded an excess return of 0.30% and 0.11% over the in-sample and out-of-sample records, respectively. The beta value was greater than 1 for the in-sample records indicating the volatility of the portfolio was greater than the benchmark volatility of the NIFTY 50 index. However, the volatility of the portfolio was lower than the benchmark for the out-of-sample records as implied by the beta value of 0.96.

**TABLE 2.11.** THE SUMMARY OF THE PERFORMANCES OF THE PORTFOLIOS ON THE IN-SAMPLE DATA
(PERIOD: JANUARY 4, 2021 – JUNE 30, 2022)

| | Auto | Bank | CD | FMCG | IT | Metal | Pharma | Pvt Bank | PSU Bank | Realty |
|---|---|---|---|---|---|---|---|---|---|---|
| Annual return | 27.05% | 5.89% | 9.51% | 11.26% | 16.12% | 42.91% | -5.00% | -12.11% | 11.02% | 38.40% |
| Cumulative return | 42.12% | 8.76% | 14.27% | 16.96% | 24.54% | 68.92% | -7.25% | -17.27% | 16.58% | 47.05% |
| Annual volatility | 24.28% | 26.30% | 18.97% | 16.00% | 26.24% | 34.28% | 19.01% | 26.12% | 37.97% | 32.65% |
| Max drawdown | -23.95% | -23.68% | -28.26% | -18.38% | -38.32% | -30.26% | -25.85% | -31.67% | -29.25% | -27.67% |
| Sharpe ratio | 1.11 | 0.35 | 0.57 | 0.75 | 0.70 | 1.21 | -0.17 | -0.36 | 0.46 | 1.16 |
| Calmar ratio | 1.13 | 0.25 | 0.34 | 0.61 | 0.42 | 1.42 | -0.19 | -0.38 | 0.38 | 1.39 |
| Sortino ratio | 1.63 | 0.48 | 0.79 | 1.07 | 0.98 | 1.69 | -0.24 | -0.49 | 0.71 | 1.74 |
| Omega ratio | 1.21 | 1.06 | 1.10 | 1.13 | 1.12 | 1.23 | 0.97 | 0.94 | 1.09 | 1.21 |
| Tail ratio | 1.05 | 0.91 | 0.98 | 1.06 | 1.03 | 0.98 | 0.93 | 1.01 | 1.08 | 1.22 |
| Skewness | -0.15 | -0.42 | -0.53 | -0.28 | -0.34 | -0.63 | 0.02 | -0.41 | 0.33 | -0.04 |
| Kurtosis | 1.40 | 2.60 | 1.84 | 0.50 | 0.60 | 2.02 | 0.85 | 2.44 | 2.95 | 1.03 |
| Stability | 0.48 | 0.03 | 0.31 | 0.44 | 0.43 | 0.56 | 0.02 | 0.48 | 0.03 | 0.32 |
| Daily value at risk | -2.95 | -3.28% | -2.35% | -1.97% | -3.23% | -4.15% | -2.41% | -3.33% | -4.71% | -3.96% |
| Alpha | 0.20 | -0.01 | 0.05 | 0.07 | 0.10 | 0.34 | -0.08 | -0.18 | 0.08 | 0.30 |
| Beta | 1.03 | 1.23 | 0.78 | 0.60 | 0.95 | 1.18 | 0.61 | 1.22 | 1.18 | 1.27 |

**TABLE 2.12.** THE SUMMARY OF THE PERFORMANCES OF THE PORTFOLIOS ON THE OUT-SAMPLE DATA
(PERIOD: JULY 1, 2022 – SEPTEMBER 20, 2023)

| | Auto | Bank | CD | FMCG | IT | Metal | Pharma | Pvt Bank | PSU Bank | Realty |
|---|---|---|---|---|---|---|---|---|---|---|
| Annual return | 37.66% | 46.36% | 16.92% | 31.80% | 23.03% | 53.18% | 31.35% | 40.56% | 104.95% | 30.08% |
| Cumulative return | 46.86% | 58.09% | 20.68% | 39.37% | 28.30% | 66.71% | 38.66% | 50.59% | 136.99% | 37.05% |
| Annual volatility | 15.04% | 17.02% | 19.44% | 12.81% | 20.37% | 21.10% | 14.18% | 17.97% | 30.90% | 19.06% |
| Max drawdown | -12.34% | -13.96% | -15.67% | -8.35% | -14.14% | -14.44% | -14.38% | -17.40% | -28.30% | -29.74% |
| Sharpe ratio | 2.20 | 2.32 | 0.90 | 2.22 | 1.12 | 2.13 | 2.00 | 1.99 | 2.48 | 1.48 |
| Calmar ratio | 3.05 | 3.32 | 1.08 | 3.81 | 1.63 | 3.68 | 2.18 | 2.33 | 3.71 | 1.01 |
| Sortino ratio | 3.46 | 3.59 | 1.36 | 3.56 | 1.67 | 3.29 | 3.40 | 3.01 | 4.09 | 2.15 |
| Omega ratio | 1.44 | 1.49 | 1.23 | 1.44 | 1.21 | 1.42 | 1.39 | 1.39 | 1.54 | 1.27 |
| Tail ratio | 1.21 | 1.21 | 1.07 | 1.33 | 1.15 | 1.03 | 1.29 | 1.06 | 1.34 | 1.06 |
| Skewness | -0.10 | -0.19 | 1.26 | -0.02 | -0.08 | -0.13 | 0.55 | -0.17 | 0.21 | -0.32 |
| Kurtosis | 1.08 | 1.35 | 53.78 | 1.10 | 0.95 | 0.81 | 0.77 | 1.07 | 2.28 | 1.08 |
| Stability | 0.65 | 0.73 | 0.00 | 0.84 | 0.50 | 0.62 | 0.46 | 0.52 | 0.70 | 0.12 |
| Daily value at risk | -1.76 | -1.99% | -2.38% | -1.50% | -2.48% | -2.48% | -1.67% | -2.12% | -3.59% | -2.29% |
| Alpha | 0.18 | 0.19 | 0.05 | 0.18 | 0.01 | 0.27 | 0.22 | 0.14 | 0.66 | 0.11 |
| Beta | 0.84 | 1.15 | 0.63 | 0.61 | 1.15 | 1.14 | 0.44 | 1.16 | 1.35 | 0.96 |

Table 2.11 presents the summary of the results of the performances of the portfolios on the in-sample data. The results of the same portfolios on the out-of-sample data are depicted in Table 2.12. Table 2.13 presents the performances of the portfolios on the overall data.

**TABLE 2.13.** THE SUMMARY OF THE PERFORMANCES OF THE PORTFOLIOS ON THE OVERALL DATA
(PERIOD: JANUARY 4, 2021 – SEPTEMBER 20, 2023)

|  | Auto | Bank | CD | FMCG | IT | Metal | Pharma | Pvt Bank | PSU Bank | Realty |
|---|---|---|---|---|---|---|---|---|---|---|
| Annual return | 31.72% | 22.50% | 12.79% | 20.08% | 19.18% | 47.44% | 9.89% | 8.58% | 46.31% | 34.16% |
| Cumulative return | 108.73% | 71.94% | 37.90% | 63.02% | 59.79% | 181.60% | 28.60% | 24.59% | 176.28% | 101.53% |
| Annual volatility | 20.63% | 22.60% | 19.17% | 14.65% | 23.76% | 29.09% | 17.03% | 22.84% | 34.98% | 26.68% |
| Max drawdown | -23.95% | -23.68% | -28.26% | -18.38% | -39.09% | -30.26% | -25.85% | -31.67% | -29.25% | -32.12% |
| Sharpe ratio | 1.44 | 1.01 | 0.72 | 1.32 | 0.86 | 1.48 | 0.64 | 0.48 | 1.26 | 1.24 |
| Calmar ratio | 1.32 | 0.95 | 0.45 | 1.09 | 0.49 | 1.57 | 0.38 | 0.27 | 1.58 | 1.06 |
| Sortino ratio | 2.15 | 1.43 | 1.03 | 1.96 | 1.23 | 2.11 | 0.94 | 0.66 | 1.98 | 1.84 |
| Omega ratio | 1.28 | 1.19 | 1.15 | 1.24 | 1.15 | 1.29 | 1.11 | 1.09 | 1.25 | 1.24 |
| Tail ratio | 0.98 | 1.00 | 0.95 | 1.08 | 1.04 | 1.18 | 1.10 | 0.99 | 1.19 | 1.09 |
| Skewness | -0.16 | -0.45 | 0.31 | -0.23 | -0.28 | -0.59 | 0.11 | -0.44 | 0.27 | -0.07 |
| Kurtosis | 2.16 | 3.31 | 26.51 | 0.84 | 0.90 | 2.80 | 1.15 | 2.89 | 2.96 | 2.01 |
| Stability | 0.87 | 0.70 | 0.04 | 0.87 | 0.03 | 0.73 | 0.00 | 0.09 | 0.66 | 0.24 |
| Daily value at risk | -2.48 | -2.76% | -2.36% | -1.77% | -2.91% | -2.91% | -2.10% | -2.84% | -4.23% | -3.23% |
| Alpha | 0.18 | 0.07 | 0.04 | 0.12 | 0.07 | 0.30 | 0.03 | -0.05 | 0.32 | 0.18 |
| Beta | 0.98 | 1.21 | 0.74 | 0.60 | 1.00 | 1.17 | 0.57 | 1.21 | 1.22 | 1.17 |

Some important observations found in the results presented in Tables 2.11, 2.12, and 2.13 are discussed in the following.

First, on the in-sample data, the rebalanced portfolio of the metal sector exhibited the best performance. The *metal* sector portfolio yielded the best values for 7 metrics out of 15 metrics used in the evaluation of the portfolios. The metrics for which the rebalanced portfolio of the *metal* sector yielded the best results are the following (i) annual return, (ii) cumulative return, (iii) Sharpe ratio, (iv) Calmar ratio, (v) Omega ratio, (vi) stability, and (vii) alpha. The FMCG sector portfolio yielded the best values for four metrics for the in-sample records. These four metrics are (i) annual volatility, (ii) max drawdown, (iii) daily value at risk, and (iv) beta. The *realty* sector portfolio yielded the best values for two metrics (i) Sortino ratio and (ii) Tail ratio. While the *PSU banks* sector portfolio produced the best value for kurtosis, the best value for skewness was yielded by the pharma sector portfolio. Overall, the performance of the *metal* sector portfolio is found to be the best on the in-sample records.

Second, on the out-of-sample records, the *PSU banks* sector portfolio is found to have outperformed all other portfolios. The 8 metrics on which the PSU banks sector portfolio yielded the best values are the following (i) annual return, (ii) cumulative return, (iii) Sharpe ratio, (iv) Sortino ratio, (v) Omega ratio, (vi) Tail ratio, (vii) kurtosis, and (viii) alpha. The FMCG sector portfolio yielded the best values for 6 metrics on the out-of-sample records. These metrics are (i) annual volatility, (ii) maximum drawdown, (iii) Calmar ratio, (iv) skewness, (v) stability, and (vi) daily value at risk. The sole metric for which the pharma sector portfolio yielded the best results is beta.

Third, on the overall data (i.e., including both in-sample and out-of-sample records), three portfolios yielded the best results for four different metrics. These three portfolios are (i) FMCG, (ii) metal, and (iii) PSU banks. The metrics for which the FMCG sector portfolio yielded the best results are (i) annual volatility, (ii) maximum drawdown, (iii) stability, and (iv) daily value at risk. The metal sector portfolio produced the best results over the following metrics on the overall data (i) annual return, (ii) cumulative return, (iii) Sharpe ratio, and (iv) Omega ratio. The four metrics for which the PSU banks sector portfolio yielded the best results are (i) Calmar ratio, (ii) Tail ratio, (iii) kurtosis, and (iv) alpha. The two metrics for which the auto sector portfolio yielded the best results are (i) Sortino ratio and (ii) stability. While the pharma sector portfolio yielded the best value of beta, the best value for kurtosis is yielded by the realty sector portfolio.

Fourth, since the performance of a portfolio on the out-of-sample records reflects the most recent performance, the rebalanced portfolios of the PSU banks sector and the FMCG sector metal are the best-performing portfolios currently. The metal sector and banking sector also appear to be quite good for investment considering their high annual and cumulative returns on the out-of-sample records.

Finally, the results indicate that the proposed rebalancing approach to portfolio design is quite effective as most of the sectors have yielded quite good returns while minimizing the risks associated with them.

Figures 2.41 to 2.56 graphically present the performance of the portfolios. The following abbreviations are used in these figures: AR: annual return, CR: cumulative return, AV: annual volatility, SR: Sharpe ratio, CR: Calmar ratio, SoR: Sortino ratio, OR: Omega ratio, SK: skewness, KU: kurtosis, ST: stability, DVaR: daily value at risk, AL: alpha, and BE: beta.

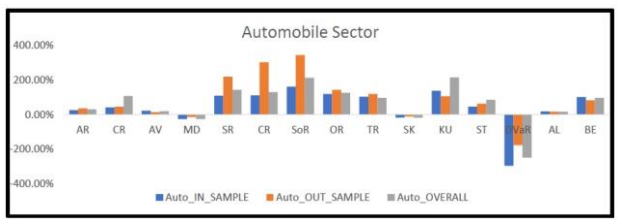

**Figure 2.41.** A graphical representation of the performance of the *auto* sector portfolio on the in-sample, out-of-sample, and overall data

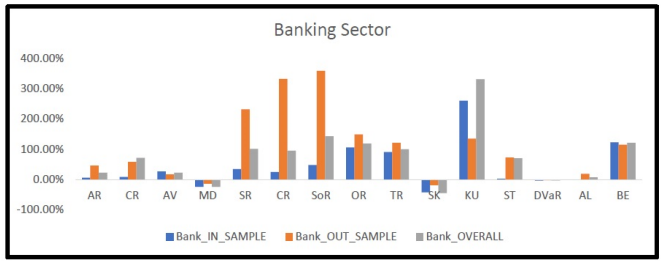

**Figure 2.42.** A graphical representation of the performance of the *banking* sector portfolio on the in-sample, out-of-sample, and overall data.

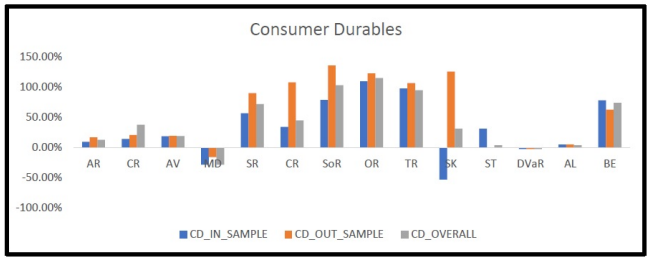

**Figure 2.43.** A graphical representation of the performance of the *consumer durables* sector portfolio on the in-sample, out-of-sample, and overall data.

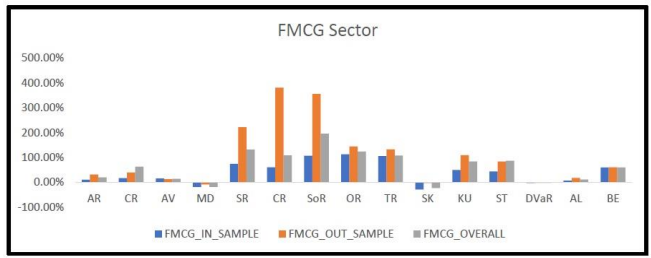

**Figure 2.44.** A graphical representation of the performance of the FMCG sector portfolio on the in-sample, out-of-sample, and overall data.

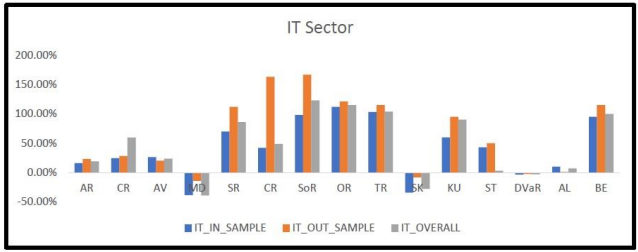

**Figure 2.45.** A graphical representation of the performance of the IT sector portfolio on the in-sample, out-of-sample, and overall data.

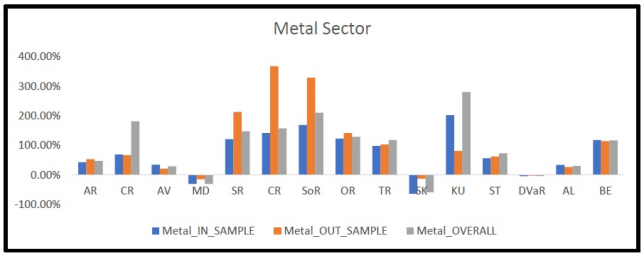

**Figure 2.46.** A graphical representation of the performance of the *metal* sector portfolio on the in-sample, out-of-sample, and overall data.

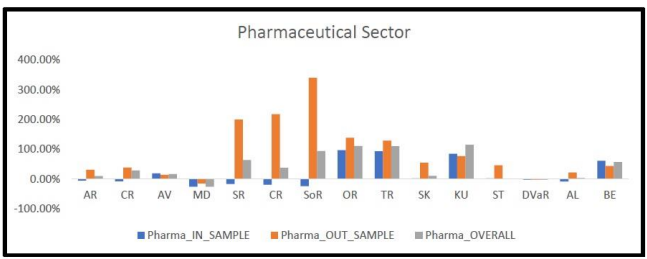

**Figure 2.47.** A graphical representation of the performance of the *pharma* sector portfolio on the in-sample, out-of-sample, and overall data.

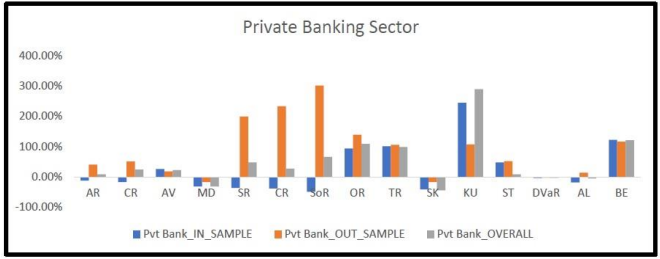

**Figure 2.48.** A graphical representation of the performance of the *private banks* sector portfolio on the in-sample, out-of-sample, and overall data.

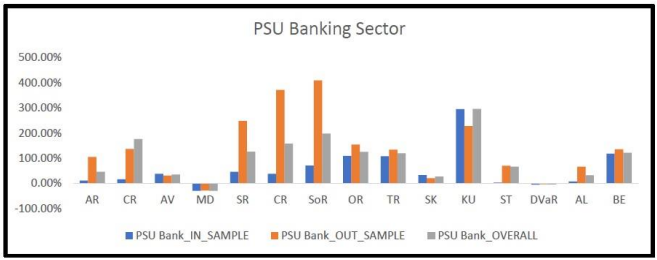

**Figure 2.49.** A graphical representation of the performance of the *PSU banks* sector portfolio on the in-sample, out-of-sample, and overall data.

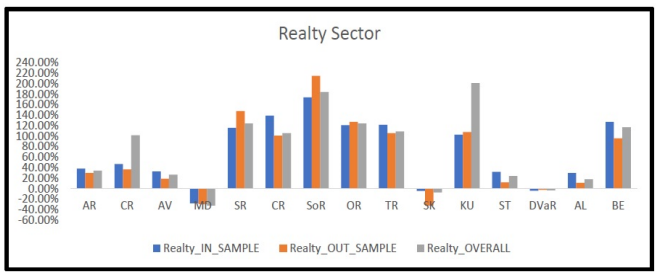

**Figure 2.50.** A graphical representation of the performance of the *realty* sector portfolio on the in-sample, out-of-sample, and overall data.

# Conclusion

This work presented in this chapter introduces a strategy for rebalancing portfolios consisting of Indian stock market investments. Initially, ten key sectors listed on the NSE from the Indian economy are selected. Within each of these sectors, the top ten stocks are identified based on their free-float market capitalization values. Sector-specific portfolios are constructed using the ten stocks for analyzing the effectiveness of the proposed rebalancing approach on these portfolios. The study utilizes historical stock prices spanning from January 4, 2021, to September 20, 2023, sourced from the NSE Website. Portfolios are created using data from January 4, 2021, to June 30, 2022, for training purposes. The portfolio performances are evaluated on the out-of-sample data from July 1, 2022, to September 20, 2023. The calendar-based rebalancing method presented in this chapter is originally set for annual rebalancing but is adaptable for weekly or monthly adjustments. The rebalanced portfolios for all ten sectors are extensively evaluated for their performance on 15 different metrics. The performance results not only provide insights into how sectors performed during the training and test periods but also gauge the overall effectiveness of the proposed portfolio rebalancing approach.

On the in-sample records from January 4, 2021, to June 30, 2022, the *metal* sector portfolio is found to have yielded the best results. The *metal* sector portfolio produced the best results for 7 metrics out of the 15 metrics used in the evaluation framework.

On the out-of-sample records from July 1, 2022, to September 20, 2023, the *PSU banks* sector portfolio yielded the best results for 8 metrics out of 15 metrics for portfolio evaluation used in the work.

In the overall analysis, which considers both in-sample and out-of-sample data from January 4, 2021, to September 20, 2023, three portfolios stood out by delivering the top results across four distinct metrics. These portfolios are (i) FMCG, (ii) *metal*, and (iii) *PSU banks*.

Because the performance on out-of-sample records reflects the most recent performance, it is worth noting that currently, the rebalanced portfolios in the *PSU banks* sector and the FMCG sector are the top-performing portfolios. Additionally, the *metal* sector and *banking* sector portfolios appear to be favorable investment options due to their impressive annual and cumulative returns on the out-of-sample records.

In conclusion, the results demonstrate that the proposed portfolio rebalancing approach is highly effective and efficient. Most sectors have generated strong returns while effectively mitigating associated risks, underscoring the robustness of this approach to portfolio rebalancing.

Future work will encompass a study involving diversified portfolios from the Indian stock market and important stocks listed on other major global stock exchanges to evaluate the effectiveness of the proposed rebalancing approach on those portfolios.

# References

Albertazzi, U., Becker, B., and Boucinha, M. (2021) "Portfolio rebalancing and the transmission of large-scale asset purchase programs: Evidence from the Euro area," *Journal of Financial Intermediation*, Volume 48, Art Id: 100896, 2021.
DOI: 10.1016/j.jfi.2020.100896.

Bernoussi, El, Rockinger, R. M (2023). "Rebalancing with transaction costs: Theory, simulations, and actual data," *Financial Market and Portfolio Management*, Vol 37, pp. 121–160, June 2023. DOI: 10.1007/s11408-022-00419-6

Chaweewanchon, A. and R. Chaysiri, R. (2022) "Portfolio optimization and rebalancing with transaction cost: A case study in the stock exchange of Thailand," *Proceedings of the 17th International Joint Symposium on Artificial Intelligence and Natural Language Processing (iSAI-NLP)*, Chiang Mai, Thailand, 2022, pp. 1-6.
DOI: 10.1109/iSAI-NLP56921.2022.9960260.

Cuthbertson, K., Hayley, S., Motson, N., and Nitzsche, D. (2016) "What does rebalancing really achieve?" *International Journal of Finance & Economics*, Vol 21, No 3, pp 224-240, February 2016. DOI: 10.1002/ijfe.1545.

Darapaneni, N., Basu, A., Savla, S., Gururajan, R., Saquib, N., Singhavi, S., Kale, A., Bid, P., and Paduri, A. R. (2020)

"Automated portfolio rebalancing using Q-learning," *Proceedings of the 11th IEEE Annual Ubiquitous Computing, Electronics & Mobile Communication Conference (UEMCON)*, New York, NY, USA, 2020, pp. 0596-0602.
DOI: 10.1109/UEMCON51285.2020.9298035.

Delpini, D., Battiston, S., Caldarelli, G., and Riccaboni, M. (2020) "Portfolio diversification, differentiation and the robustness of holdings networks," *Applied Network Science*, Vol 5, Art Id: 37, July 2020. DOI: 10.1007/s41109-020-00278-y.

Fischer, A. M., Greminger, R. P., Grisse, C., and Kaufmann, S. "Portfolio rebalancing in times of stress," *Journal of International Money and Finance*, Vol 113, Art Id: 102360, May 2021. DOI: 10.1016/j.jimonfin.2021.102360.

Guo, X. and Ryan, S. M. (2023) "Portfolio rebalancing based on time series momentum and downside risk," *IMA Journal of Management Mathematics*, Vol 34, No 2, pp. 355–381, April 2023. DOI: 10.1093/imaman/dpab037.

Guo, X. and Ryan, S. M. (2021) "Portfolio rebalancing based on time series momentum and downside risk," *IMA Journal of Management Mathematics*, Vol 34, No 2, pp 355-381, November 2021. DOI: 10.1093/imaman/dpab037.

Hagiwara, D. and Harada, T. (2017) "Portfolio Rebalancing with a consideration of market conditions changes using instance-based policy optimization," *Electronics and Communications in Japan*, Vol 100, No 4, pp. 66-75, April 2017.
DOI: 10.1002/ecj.11950.

Hau, H. and Rey, H. (2004) "Can portfolio rebalancing explain the dynamics of equity returns, equity flows, and exchange rates?" *American Economic Review*, Vol 94, No 2, pp. 126-133, May 2004. DOI: 10.1257/0002828041302389.

Hilliard, J.E., Hilliard, J. (2018) "Rebalancing versus buy and hold: theory, simulation and empirical analysis," *Review of Quantitative Finance and Accounting*, Vol 50, pp. 1–32, 2018. DOI: 10.1007/s11156-017-0621-5.

Horn, M., Oehler, A. (2020) "Automated portfolio rebalancing: Automatic erosion of investment performance?" *Journal of Asset Management*, Vol 21, pp. 489–505, 2020.

DOI: 10.1057/s41260-020-00183-0.

Jaconetti, C. M., Kinniry, F. M., and Zilbering (2010) "Best practices for portfolio rebalancing", *Vanguard Research*, Valley Forge, PA, USA, July 2010.

Ji, R., Lejeune, M. A., and Prasad, S. Y. (2017) "Dynamic portfolio optimization with risk-aversion adjustment utilizing technical indicators," *Proceedings of the 20th International Conference on Information Fusion (Fusion)*, Xi'an, China, 2017, pp. 1-8. DOI: 10.23919/ICIF.2017.8009871.

Jiang, Z., Ji, R., and Chang, K-C. (2020) "A machine learning integrated portfolio rebalance framework with risk-aversion adjustment," *Journal of Risk and Financial Management*, Vol 13, No 7, Art Id: 155, July 2020. DOI: 10.3390/jrfm13070155.

Kim, K. and Lee, D. (2020). "Equity market integration and portfolio rebalancing," *Journal of Banking & Finance*, Vol 113, Art Id: 105775, April 2020. DOI: 10.1016/j.jbankfin.2020.105775.

Kutner, M., Nachtsheim, C., and Neter, J. (2004) *Applied Linear Regression Models*, 4th Edition, McGraw-Hill Education. ISBN-13: 978-0073014661.

Laher, S., Paskaramoorthy, A., and Van Zyl, T. L. (2021) "Deep learning for financial time series forecast fusion and optimal portfolio rebalancing," *Proceedings of the IEEE 24th International Conference on Information Fusion (FUSION'21)*, Sun City, South Africa, 2021, pp. 1-8. DOI: 10.23919/FUSION49465.2021.9626945.

Maree, C. and Omlin, C. W. (2022) "Balancing profit, risk, and sustainability for portfolio management," *Proceedings of the 2022 IEEE Symposium on Computational Intelligence for Financial Engineering and Economics (CIFEr)*, Helsinki, Finland, 2022, pp. 1-8. DOI: 10.1109/CIFEr52523.2022.9776048.

Markowitz, H. (1952) "Portfolio selection," *The Journal of Finance*, Vol 7, No 1, pp. 77-91. DOI: 10.2307/2975974.

Mehtab S. and Sen, J. (2022) "Analysis and forecasting of financial time series using CNN and LSTM-based deep learning models," in: Sahoo, J. P., Tripathy, A. K., Mohanty, M., Li, K. C., Nayak, A, K. (eds), *Advances in Distributed Computing and Machine*

*Learning*, Lecture Notes in Networks and Systems, Vol 302, pp. 405-423, Springer, Singapore.
DOI: 10.1007/978-981-16-4807-6_39.

Mehtab, S. and Sen, J. (2021) "A time series analysis-based stock price prediction using machine learning and deep learning models," *International Journal of Business Forecasting and Marketing Intelligence*, Vol 6, No 4, pp. 272-335.
DOI: 10.1504/IJBFMI.2020.115691.

Mehtab, S. and Sen, J. (2020a) "Stock price prediction using convolutional neural networks on a multivariate time series," *Proceedings of the 3rd National Conference on Machine Learning and Artificial Intelligence (NCMLAI'20)*, February 1, 2020, New Delhi, India. DOI: 10.36227/techrxiv.15088734.v1.

Mehtab, S. and Sen, J. (2020b) "Stock price prediction using CNN and LSTM-based deep learning models," *Proceedings of the IEEE International Conference on Decision Aid Sciences and Applications (DASA'20)*, pp. 447-453, November 8-9, 2020, Sakheer, Bahrain. DOI: 10.1109/DASA51403.2020.9317207.

Mehtab, S. and Sen, J. (2019) "A robust predictive model for stock price prediction using deep learning and natural language processing," *Proceedings of the 7[th] International Conference on Business Analytics and Intelligence (BAICONF'19)*, December 5-7, 2019, Bangalore, India.
DOI: 10.36227/techrxiv.15023361.v1.

Mehtab, S., Sen, J. and Dutta, A. (2021) "Stock price prediction using machine learning and LSTM-based deep learning models", in: Thampi, S. M., Piramuthu, S., Li, K. C., Beretti, S., Wozniak, M., Singh, D. (eds), *Machine Learning and Metaheuristics Algorithms, and Applications (SoMMA'20)*, pp 86-106, Communications in Computer and Information Science, Vol 1366, Springer, Singapore.
DOI: 10.1007/978-981-16-0419-5_8.

Mehtab, S., Sen, J. and Dasgupta, S. (2020) "Robust analysis of stock price time series using CNN and LSTM-based deep learning models," *Proceedings of the IEEE 4[th] International Conference on Electronics, Communication and Aerospace Technology*

*(ICCEA'20)*, pp. 1481-1486, November 5-7, 2020, Coimbatore, India. DOI: 10.1109/ICECA49313.2020.9297652.

NSE Website: http://www1.nseindia.com (Accessed on August 30, 2023)

NIFTY 50 Wiki page: https://en.wikipedia.org/wiki/NIFTY_50 (Accessed on August 30, 2023)

DOI: 10.2139/ssrn.3237540.

Pai, G. A. V. (2018) "Active portfolio rebalancing using multi-objective metaheuristics," *Proceedings of the 2018 IEEE Symposium Series on Computational Intelligence (SSCI)*, Bangalore, India, 2018, pp. 1845-1852.

DOI: 10.1109/SSCI.2018.8628875.

Pagliaro, C. A. and Utkus, S. P. (2019) "Assessing the value of advice," *Vanguard Research*, Valley Forge, PS, USA, September 2019.

Sen, J. (2023) "Portfolio optimization using reinforcement learning and hierarchical risk parity approach," in: Rivera, G., Cruz-Reyes, L., Dorronsoro, B., and Rosete-Suarez, A, (eds) *Data Analytics and Computational Intelligence: Novel Models, Algorithms and Applications,* Studies in Big Data, Vol 132, pp. 509-554, Springer, Cham, September 2023.

DOI: 10.1007/978-3-031-38325-0_20.

Sen, J. (2022a) "A forecasting framework for the Indian healthcare sector index," *International Journal of Business Forecasting and Marketing Intelligence* (*IJBFMI*), Vol 7, No 4, pp. 311-350, 2021. DOI: 10.1504/IJBFMI.2022.10047095.

Sen, J. (2022b) "Optimum pair-trading strategies for stocks using cointegration-based approach," *Proceedings of the IEEE 29th OITS International Conference on Information Technology (OCIT'22)*, December 14-16, 2022, Bhubaneswar, India.

DOI: 10.1109/OCIT56763.2022.00076.

Sen, J. (2022c) "Designing efficient pair-trading strategies using cointegration for the Indian stock market," *Proceedings of the IEEE 2nd Asian Conference on Innovation in Technology (ASIANCON'22)*, pp. 1 - 9, Pune, India, August 2022.

DOI: 10.1109/ASIANCON55314.2022.9909455.

Sen, J. (2022d) "A Comparative Analysis of Portfolio Optimization Using Reinforcement Learning and Hierarchical Risk Parity Approaches," *Proceedings of the 9th International Conference on Business Analytics and Intelligence (BAICONF'22), December 15-17, 2022, Bangalore, India.*

Sen, J. (2018a) "Stock price prediction using machine learning and deep learning frameworks," *Proceedings of the 6th International Conference on Business Analytics and Intelligence (ICBAI'18)*, December 20-22, Bangalore, India.

Sen, J. (2018b) "Stock composition of mutual funds and fund style: A time series decomposition approach towards testing for consistency," *International Journal of Business Forecasting and Marketing Intelligence*, Vol 4, No 3, pp. 235-292. DOI: 10.1504/IJBFMI.2018.092781.

Sen, J. (2017) "A time series analysis-based forecasting approach for the Indian realty sector," *International Journal of Applied Economic Studies*, Vol 5, No 4, pp. 8-27. DOI: 10.36227/techrxiv.16640212.v1.

Sen, J. and Datta Chaudhuri, T. (2018) "Understanding the sectors of Indian economy for portfolio choice," *International Journal of Business Forecasting and Marketing Intelligence*, Vol 4, No 2, pp. 178-222. DOI: 10.1504/IJBFMI.2018.090914.

Sen, J. and Datta Chaudhuri, T. (2017a) "A robust predictive model for stock price forecasting," *Proceedings of the 5th International Conference on Business Analytics and Intelligence (BAICONF'17)*, December 11-13, 2017, Bangalore, India. DOI: 10.36227/techrxiv.16778611.v1.

Sen, J. and Datta Chaudhuri, T. (2017b) "A predictive analysis of the Indian FMCG sector using time series decomposition-based approach," *Journal of Economics Library*, Vol 4, No 2, pp. 206-226. DOI: 10.1453/jel.v4i2.1282.

Sen, J. and Datta Chaudhuri, T. (2016a) "Decomposition of time series data to check consistency between fund style and actual fund composition of mutual funds," *Proceedings of the 4th International Conference on Business Analytics and Intelligence (ICBAI'16)*, December 19-21, 2016. DOI: 10.13140/RG.2.2.33048.19206.

Sen, J. and Datta Chaudhuri, T. (2016b) "An investigation of the structural characteristics of the Indian IT sector and the capital goods sector – An application of the R programming language in time series decomposition and forecasting," *Journal of Insurance and Financial Management*, Vol 1, No 4, pp 68-132. DOI: 10.36227/techrxiv.16640227.v1.

Sen, J. and Datta Chaudhuri, T. (2016c) "An alternative framework for time series decomposition and forecasting and its relevance for portfolio choice – A comparative study of the Indian consumer durable and small cap sectors," *Journal of Economics Library*, Vol 3, No 2, pp. 303-326. DOI: 10.1453/jel.v3i2.787.

Sen, J. and Datta Chaudhuri, T. (2016d) "Decomposition of time series data of stock markets and its implications for prediction – An application for the Indian auto sector," *Proceedings of the 2nd National Conference on Advances in Business Research and Management Practices (ABRMP'16)*, pp. 15-28, January 8-9, 2016. DOI: 10.13140/RG.2.1.3232.0241.

Sen, J. and Datta Chaudhuri, T. (2015) "A framework for predictive analysis of stock market indices – A study of the Indian auto sector," *Journal of Management Practices*, Vol 2, No 2, pp. 1-20. DOI: 10.13140/RG.2.1.2178.3448.

Sen, J. and Dutta, A. (2022a) "A comparative study of hierarchical risk parity portfolio and eigen portfolio on the NIFTY 50 stocks," in: Buyya, R., Hernandez, S.M., Kovvur, R.M.R., Sarma, T.H. (eds) *Computational Intelligence and Data Analytics,* Lecture Notes on Data Engineering and Communications Technologies, Vol 142, pp. 443-460, Springer, Singapore. DOI: 10.1007/978-981-19-3391-2_34.

Sen, J. and Dutta, A. (2022b) "Design and Analysis of Optimized Portfolios for Selected Sectors of the Indian Stock Market," *Proceedings of the 2022 International Conference on Decision Aid Sciences and Applications (DASA)*, pp. 567-573, March 23-25, 2022, Chiangrai, Thailand.
DOI: 10.1109/DASA54658.2022.9765289.

Sen, J. and Dutta, A. (2022c), "Portfolio Optimization for the Indian Stock Market," in: Wang, J. (ed.) *Encyclopaedia of Data Science and Machine Learning*, pp. 1904-1951, IGI Global, USA, August 2022. DOI: 10.4018/978-1-7998-9220-5.ch115.

Sen, J. and Dutta, A. (2021) "Risk-based portfolio optimization on some selected sectors of the Indian stock market," *Proceedings of the 2ⁿᵈ International Conference on Big Data, Machine Learning and Applications (BigDML'21)*, December 19-20, 2021, Silchar, India. (In press)

Sen, J. and Mehtab, S. (2022a) "A comparative study of optimum risk portfolio and eigen portfolio on the Indian stock market," *International Journal of Business Forecasting and Marketing Intelligence*, Vol 7, No 2, pp 143-195.
DOI: 10.1504/IJBFMI.2021.120155.

Sen, J. and Mehtab, S. (2022b) "Long-and-Short-Term Memory (LSTM) Price Prediction-Architectures and Applications in Stock Price Prediction," in: Singh, U., Murugesan, S., and Seth, A. (eds) *Emerging Computing Paradigms - Principles, Advances, and Applications*, Wiley, USA, 2022.
DOI: 10.1002/9781119813439.ch8.

Sen, J. and Mehtab, M. (2021a) "Design and analysis of robust deep learning models for stock price prediction," in: Sen, J. (ed) *Machine Learning – Algorithms, Models and Applications*, pp. 15-46, IntechOpen, London, UK.
DOI: 10.5772/intechopen.99982.

Sen J. and Mehtab, S. (2021b) "Accurate stock price forecasting using robust and optimized deep learning models," *Proceedings of the IEEE International Conference on Intelligent Technologies (CONIT)*, pp. 1-9, June 25-27, 2021, Hubballi, India.
DOI: 10.1109/CONIT51480.2021.9498565.

Sen A. and Sen, J. (2023) "A study of the performance evaluation of equal-weight portfolio and optimum risk portfolio on the Indian stock market," *International Journal of Business Forecasting and Marketing Intelligence (IJBFMI)*, Inderscience Publishers, 2023. (In Press)

Sen, J., Dutta, A. and Mehtab, S. (2021a) "Stock portfolio optimization using a deep learning LSTM model," *Proceedings of the IEEE Mysore Sub Section International Conference (MysuruCon'21)*, pp. 263-271, October 24-25, 2021, Hassan, Karnataka, India.
DOI: 10.1109/MysuruCon52639.2021.9641662.

Sen, J., Dutta, A. and Mehtab, S. (2021b) "Profitability analysis in stock investment using an LSTM-based deep learning model," *Proceedings of the IEEE 2nd International Conference for Emerging Technology (INCET'21)*, pp. 1-9, May 21-23, Belagavi, India. DOI: 10.1109/INCET51464.2021.9456385.

Sen, J., Dutta, A., Mondal, S. and Mehtab, S. (2021c) "A comparative study of portfolio optimization using optimum risk and hierarchical risk parity approaches," *Proceedings of the 8th International Conference on Business Analytics and Intelligence (ICBAI'21)*, December 20-22, Bangalore, India. DOI: 10.13140/RG.2.2.35308.28809.

Sen, J., Mehtab, S. and Dutta, A. (2021d) "Volatility modeling of stocks from selected sectors of the Indian economy using GARCH," *Proceedings of the IEEE Asian Conference on Innovation in Technology (ASIANCON'21)*, pp. 1-9, August 28-29, 2021, Pune, India.
DOI: 10.1109/ASIANCON51346.2021.9544977.

Sen, J., Mehtab, S., Dutta, A. and Mondal, S. (2021e) "Precise stock price prediction for optimized portfolio design using an LSTM model," *Proceedings of the IEEE 19th International Conference on Information Technology (OCIT'12)*, pp. 210-215, December 16-18, 2021, Bhubaneswar, India.
DOI: 10.1109/OCIT53463.2021.00050.

Sen, J., Mehtab, S., Dutta, A. and Mondal, S. (2021f) "Hierarchical risk parity and minimum variance portfolio design on NIFTY 50 stocks," *Proceedings of the IEEE International Conference on Decision Aid Sciences and Applications (DASA'21)*, December 7-8, 2021, Sakheer, Bahrain.
DOI: 10.1109/DASA53625.2021.9681925.

Sen, J., Mondal, S. and Nath, G. (2021g) "Robust portfolio design and stock price prediction using an optimized LSTM model," *Proceedings of the IEEE 18th India Council International Conference (INDICON'21)*, pp. 1-6, December 19-21, 2021, Guwahati, India.
DOI: 10.1109/INDICON52576.2021.9691583.

Sen, J., Mondal, S. and Mehtab, S. (2021h) "Portfolio optimization on NIFTY thematic sector stocks using an LSTM model,"

*Proceedings of the IEEE International Conference on Data Analytics for Business and Industry (ICDABI'21)*, pp. 364-369, October 25-26, 2021, Bahrain.

DOI: 10.1109/ICDABI53623.2021.9655886.

Sen, J., Mondal, S., and Mehtab, S. (2021i) "Analysis of Sectoral Profitability of the Indian Stock Market Using an LSTM Regression Model," *Proceedings of the Deep Learning Developers' Conference (DLDC'21)*, September 24, 2021, Bangalore, India. DOI: 10.36227/techrxiv.17048579.v1.

Sen, J., Mehtab, S. and Nath, G. (2020) "Stock price prediction using deep learning models," *Lattice: The Machine Learning Journal*, Vol 1, No 3, pp. 34-40. DOI: 10.36227/techrxiv.16640197.v1.

Strub, O. (2017) "A new MILP formulation for rebalancing enhanced index-tracking portfolios," *Proceedings of the 2017 IEEE International Conference on Industrial Engineering and Engineering Management (IEEM)*, Singapore, 2017, pp. 989-993. DOI: 10.1109/IEEM.2017.8290040.

Thakkar, A. and Chaudhuri, K. (2021) "A comprehensive survey on portfolio optimization, stock price and trend prediction using particle swarm optimization," *Archives of Computational Methods in Engineering*, Vol 28, pp. 2133-2164.

DOI: 10.1007/s11831-020-09448-8.

Tunc, S., Donmez, M. A., and Kozat, S. S. (2013) "Growth optimal investment with threshold rebalancing portfolios under transaction costs," *Proceedings of the 2013 IEEE International Conference on Acoustics, Speech and Signal Processing*, Vancouver, BC, Canada, 2013, pp. 8717-8721.

DOI: 10.1109/ICASSP.2013.6639368.